\documentclass{aa}  
\usepackage{graphicx}
\usepackage{txfonts}
\usepackage{color}
\bibpunct{(}{)}{;}{a}{}{,} 
\newcommand{\beq}{\begin{equation}}
\newcommand{\eeq}{\end{equation}}
\newcommand{\bea}{\begin{eqnarray}}
\newcommand{\eea}{\end{eqnarray}}
\newcommand{\arate}{\langle\dot{M}\rangle}
\newcommand{\alamb}{\langle\lambda\rangle}
\newcommand{\msol}{{\,\rm M}_{\odot}}

\begin{document} 

\title{Constraining supermassive black hole evolution through the
  continuity equation}

\titlerunning{SMBH and AGN evolution}

\author{Marco Tucci \inst{1} \and Marta Volonteri \inst{2}}

\institute{D\'epartement de Physique Th\'eorique and Center for
  Astroparticle Physics, Universit\'e de Gen\`eve,~~24 quai Ansermet,
  CH--1211 Gen\`eve 4, Switzerland \\ \email{Marco.Tucci@unige.ch}
  \and Institut d'Astrophysique de Paris, Sorbonne Universit\`es, UPMC
  Univ Paris 6 et CNRS, UMR 7095, 98 bis bd Arago, 75014 Paris, France \\
  \email{martav@iap.fr} }

\date{\today}
   
\abstract{The population of supermassive black holes (SMBHs) is
  split between those that are quiescent, such as those seen in local galaxies
  including the Milky Way, and those that are active, resulting in quasars and
  active galactic nuclei. Outside our neighborhood, all the
  information we have on SMBHs is derived from quasars and active
  galactic nuclei (AGN), giving us a partial view. We study the evolution of
  the SMBH population, total and active, by
  the continuity equation, backwards in time from $z=0$ to
  $z=4$. Type-1 and type-2 AGN are differentiated in our model on the
  basis of their respective Eddington ratio distributions, chosen on the basis of
  observational estimates. The duty cycle is obtained by matching the
  luminosity function of quasars, and the average radiative efficiency
  is the only free parameter in the model. For higher radiative
  efficiencies ($\gtrsim 0.07$), a large fraction of the SMBH population,
  most of them quiescent, must already be in place by $z=4$. For lower
  radiative efficiencies ($\sim0.05$), the duty cycle increases with
  the redshift and the SMBH population evolves dramatically from
  $z=4$ onwards. The mass function of active SMBHs does not depend on the
  choice of the radiative efficiency or of the local SMBH mass
  function, but it is mainly determined by the quasar luminosity
  function once the Eddington ratio distribution is fixed. Only   direct measurement of the total black-hole mass function (BHMF) at redshifts $z\gtrsim 2$ could
  break these degeneracies, offering important constraints on the average
  radiative efficiency. Focusing on type-1 AGN, for which
  observational estimates of the mass function and Eddington ratio
  distribution exist at various redshifts, models with lower radiative
  efficiencies better reproduce the high-mass end of the mass function
  at high z, but tend to over-predict it at low z, and vice-versa
  for models with higher radiative efficiencies.  }


   \maketitle
%

\section{Introduction}
\label{sec1}
The population of supermassive black holes (SMBHs) we detect  at the center of local galaxies \citep[e.g.,][]{1995ARA&A..33..581K} has been built over long cosmic times, as traced, for instance, by the luminosity function of active galactic nuclei (AGN) and quasars. Local SMBHs represent the relics of the young population observed at earlier cosmic epochs. 

While at $z=0$, for at least several tens of galaxies, we can measure the SMBH mass  \citep[for a review of the masses and the measurement techniques, we refer to][]{kor13}; this is impossible at higher redshift, where we can only obtain estimates of SMBH masses through indirect methods. In fact, as soon as we move away from the local Universe, we can only observe active SMBHs, powering AGN and quasars, giving us only a partial view of the full population. For each AGN, there may be several dormant SMBHs that we cannot identify as such. 

There are different ways of theoretically estimating the cosmic
evolution of the SMBH population. Analytical models, for instance, can
use the continuity equation \citep[see,
e.g.,][]{cav71,sma92,yu02,mar04,mer08,sha13}, which relies on the
assumption that the mass function of SMBHs at one time can be evolved
in time, backwards or forwards, to predict the mass function at an
earlier or later time, depending on the distribution of
accretion rates. This kind of models assumes that SMBHs grow
primarily by gas accretion in luminous phases, and that the luminosity
function of AGN can be used as a constraint. Another option is to
convolve the dark matter halo mass function and/or merger rate with a
relation between SMBH and halo, and track the SMBHs by assuming that
the link with the properties of dark matter halos is sufficient to
describe the main properties of the SMBH population
\citep[e.g.,][]{1998ApJ...503..505H,2002ApJ...581..886W,2003ApJ...595..614W}.

Alternatively, one can use semi-analytical models, where the
population of SMBHs is evolved jointly with the dark matter halos and
galaxies hosting them, using analytical prescription to describe the
main physical processes such as gas cooling, star-formation rate, and
accretion onto the SMBHs
\citep[e.g.,][]{Kauffmann2000,Cattaneo2001,VHM,2000MNRAS.311..279M,2000MNRAS.318L..35H,Cattaneo2005,Croton2006,
  2006MNRAS.370..645B,2007MNRAS.375.1189M,Fontanot2011,Hirschmann2012}. The
main assumption in this case is that the properties of dark
matter halos determine those of the galaxies and SMBHs, and that
baryonic physics can be described with analytical expressions.

Finally, one can use cosmological hydrodynamical simulations \citep[e.g.,][]{Sijacki2007, Dimatteo2008, Booth2009,duboisetal10,2012MNRAS.420.2662D,2014MNRAS.442.2304H,2015MNRAS.452..575S,2016arXiv160201941V}, where the advantage is that the cosmological environment is followed faithfully in the non-linear regime, and the growth of galaxies through mergers and gas accretion is naturally taken into account. However, small-scale baryonic physics is included using analytical approximations as in the case of semi-analytical models, and such simulations are computationally expensive, excluding the possibility of exploring a large parameter space, as is instead possible using analytical or semi-analytical techniques.  

This paper is the first in a series of two, where we first develop a framework based on the continuity equation to follow the cosmic evolution of SMBHs (this paper) and we then apply this framework to model the AGN luminosity function at radio wavelengths, where little dedicated work has thus far been done \citep[e.g.,][]{2004ApJ...612..698H,2010MNRAS.401.1869S}.

The outline of the paper is as follows. In Section 2, we recall the main properties of the continuity equation and its implementation. In Section 3, we describe our methodology and assumptions. In Section 4, we present the main results and their discussion, and, finally,  in Section 5, we summarize our conclusions.

\section{SMBH evolution via a continuity equation}
\label{sec2}

The evolution of the SMBH population through
 time is typically described by a continuity equation \citep[see,
e.g.,][]{cav71,sma92,yu02,mar04,mer08,sha13}:
\beq
\frac{\partial\Phi_{BH}}{\partial t}(M,t)=-M\,\frac{\partial}{\partial M}
\Bigg[\frac{\arate(M,t)\,\Phi_{BH}(M,t)}{M}\Bigg]\,
\label{s2e1}
,\eeq
where $\Phi_{BH}(M,t)$ (hereafter, we use the symbol
  $\Phi(x)$ to denote functions in logarithmic units,
  that is, $\Phi(x)\equiv dN/d\log(x)$) is the black-hole mass function
(BHMF), defined as the number of SMBHs per co-moving volume with mass
in the logarithmic interval $\log(M),\,\log(M)+\Delta\log(M)$ and
$\arate$ is the average accretion rate of SMBHs of mass $M$ at time
$t$. Under the assumption that SMBHs grow during phases of AGN
activity, the growth rate is the result of the mass that falls into
the black hole but is not converted into energy. If $\dot{M}_{acc}$ is
the accretion rate of matter onto a SMBH, the part of it converted into
luminosity is $L=\epsilon\dot{M}_{acc}c^2$, where $\epsilon$ is the
radiative efficiency of the accretion flow. The growth rate of a SMBH
is thus given by $\dot{M}=(1-\epsilon)\dot{M}_{acc}$. Consequently,
the bolometric luminosity can be related to the mass accretion rate by
\beq
L=\frac{\epsilon}{(1-\epsilon)}c^2\dot{M}(M,t)\,.
\label{s2e2}
\eeq
Although individual SMBHs turn on and off, the mass function evolution
depends only on the average accretion rate of all SMBHs, active and
inactive. In terms of the Eddington ratio, defined as $\lambda\equiv
L/L_{\rm Edd}$ with the Eddington luminosity $L_{\rm Edd}=\ell M$ and
$\ell\simeq1.26\times10^{38}\,\msol^{-1}$\,erg\,s$^{-1}$, the
average accretion rate is
\beq
\frac{\arate(M,t)}{M}=\frac{(1-\epsilon)\ell} {\epsilon c^2}\,U(M,t)\,\alamb(M,t)\,,
\label{s2e3}
\eeq
where $\epsilon$  now has to be considered as the average value for
SMBHs of mass $M$ at time $t$. The average Eddington ratio
$\alamb$ is
\beq
\alamb(M,t)=\int d\log \lambda\,P(\lambda|M,t)\,\lambda\,,
\label{s2e4}
\eeq
where $P(\lambda|M,t)$ is the Eddington ratio distribution, that is, the
probability for a SMBH of mass $M$ to accrete at the Eddington ratio
$\lambda$ at time $t$ (it is per unit of $\log \lambda$ and is
normalized to unity, i.e., $\int d\log \lambda\,P(\lambda|M,t)=1$).

The average accretion rate also depends on the duty cycle, $U(M,t)$,
that is, on the fraction of SMBHs of mass $M$ that are active at time
$t$. By the definition of duty cycle, we can introduce the mass
function of active SMBHs (throughout the paper, this is also referred to
as the AGN MF):
\beq
\Phi_{AGN}(M,t)=\Phi_{BH}(M,t)\,U(M,t).
\label{s2e5}
\eeq
Using Eqs.\,\ref{s2e3} and \ref{s2e5}, the continuity equation can be
written as
\beq
\frac{\partial\Phi_{BH}}{\partial z}=
-\frac{\ell}{c^2\ln(10)}\,\frac{dt}{dz}\frac{\partial}{\partial\log M}
\Bigg[\frac{(1-\epsilon)}{\epsilon}\alamb\Phi_{AGN}\Bigg]\,.
\label{s2e6}
\eeq
The AGN MF can be also related to the bolometric luminosity function
of quasars by the following equation
\beq
\Phi(L,z)=\int d\log\lambda\,P(\lambda|M,z)\,\Phi_{AGN}(M,z)\,.
\label{s2e7}
\eeq
The quasar luminosity function is usually taken as input in models
based on the continuity equation, and used to constrain the evolution
of the AGN MF.

Finally, we note that the contribution from black hole mergers has not
been included in Eq.\,\ref{s2e1}. The relevance of this term is relatively
unknown and is difficult to evaluate. \citet{sha13} attempted to
estimate this effect using the SMBH merger rate predicted by
hierarchical models of structure formation. They found that mergers
have limited effects at $z>2$, but that they might be relevant on the local
mass function increasing the space density of high--mass SMBHs
\citep[see also the discussion in][]{mer08}. \citet{ave15} agree with
these conclusions. They found that mergers moderately increase the
space densities of SMBHs with $M\ga10^9\msol$ at low redshifts, but
that the effect is probably smaller than the current uncertainties on
the local BHMF. Numerical simulations by
\citet{vol05,ber08} also verified that mass accretion dominates over
mergers in determining the mass growth and spin distribution of black
holes. Because of the large uncertainties in both merging rates of
galaxies and physical processes involved in black hole mergers, we
neglect this contribution in the following analysis. This is a
standard assumption in most of the studies that employ the continuity
equation.

\section{Method}
\label{sec3}

Below, we describe the method we employ to solve the continuity
equation (Eq.\,\ref{s2e6}). We use the local
BHMF as boundary condition and integrate the equation backwards in time up to
redshift $z=4$. The continuity equation also requires the simultaneous
knowledge of the Eddington ratio distribution, the duty cycle, and the
average radiative efficiency of SMBHs.

Briefly, the inputs of the model are the following:
\begin{itemize}

\item {\it Local BHMF}. We use two possible mass functions, based
  on results of \citet{sha09} and \citet{sha13b}.

\item {\it Eddington ratio distribution}. We consider different
  distributions for type-1 and type-2 AGN on the basis of the
  observational estimates of \citet{kel13} and \citet{air12}.

\item {\it Average radiative efficiency $\epsilon$}. This is a free
  parameter of the model. We study solutions of the continuity
  equation for different values of $\epsilon$, assuming it to be
  independent of time and of SMBH mass.

\item {\it Quasar luminosity function (QLF)}. We adopt the
  observational QLF from \citet{hop07}.

\end{itemize}
Finally, the duty cycle is described by a parametric function (double
power law) of redshift and SMBH mass. The parameters are determined as
the best fit of the observational quasar luminosity function,
employing a Markov Chain Monte Carlo (MCMC) method. Jointly with the
BHMF, the duty cycle is then the main output of the model.

In the following, we detail the main ingredients and assumptions of
the model.

\subsection{Local BHMF}
\label{s3s1}

\begin{figure}
  \centering
  \includegraphics[width=\hsize]{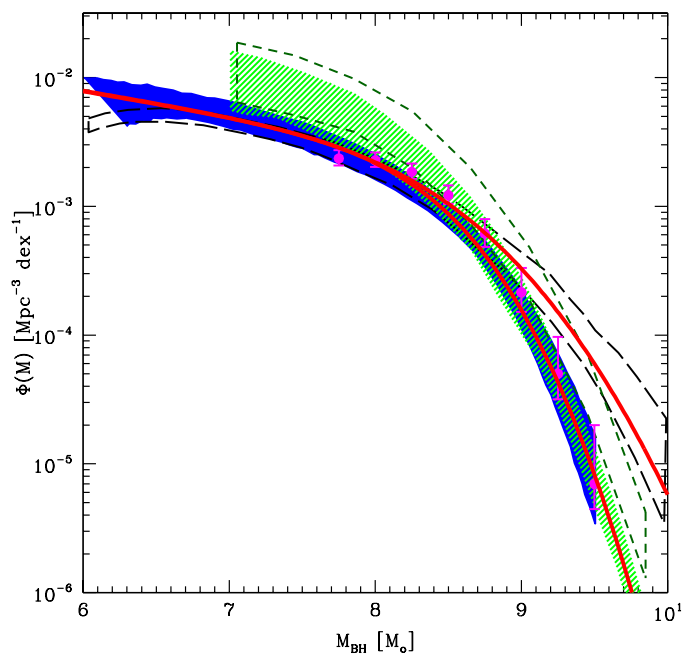}
  \caption{Local BHMF as estimated by \citet{sha09} (blue shaded
    area), \citet{vik09} (magenta points), \citet{li11} (green lined
    area), \citet{sha13b} (black dashed region), and \citet{ued14}
    (dark green short--dashed region). Red solid thick lines
    correspond to the Schecter function convolved with Gaussian
    scatter of 0.3\,dex (lower line) and 0.5\,dex (upper line).}
  \label{s3f1}
\end{figure}

Several works estimated the local BHMF using the observed empirical
scaling relations between SMBH mass and host galaxy spheroidal
properties, such as luminosity, stellar velocity dispersion, and bulge
mass \citep[see][for a review]{sha09,kor13}. \citet{sha09} present a
compilation of local BHMF determinations based on a variety of
methods, scaling relations, and data sets. The spread in the estimates
is shown in Fig.\,\ref{s3f1} (the blue shaded area). These estimates
give a local mass density of SMBHs in the range
$\rho_{BH}=3.2$--$5.4\times10^5\,\msol$Mpc$^{-3}$. Moreover,
\citet{vik09} derive the local BHMF for a large sample of galaxies
from the Millennium Galaxy Catalogue using the empirical black hole
mass--bulge luminosity relation of \citet{gra07}. \citet{li11} use
luminosity and stellar mass functions of field galaxies to constrain
the masses of their spheroids, and then compute SMBH mass through the
empirical correlation between SMBH and spheroid mass
\citep{har04}. Despite the different methods, all these derived local
BHMFs are consistent with one another (see Fig.\,\ref{s3f1}).

However, uncertainties on the actual local BHMF do remain. The main
issue is related to the scaling relationships used for the SMBH mass
determination. Recently, \citet{kor13} have updated the calibration
between the SMBH mass and the luminosity, mass, or velocity dispersion
of the bulge component of the host galaxy in the local universe. This
led to an upward revision by a factor of approximately 2-3 of previously computed SMBH masses. Assuming the new relation of
\citet{kor13}, \citet{ued14} updated the local mass function of
\citet{li11} finding a larger MF at all masses. Moreover,
\citet{sha13b} derived the local BHMF from the assumption that all
local galaxies follow the revised early-type $M_{BH}$--$\sigma$
relation from \citet{mcc13}. In this case, the revised estimates give
a significantly larger local MF at the highest masses ($\ga10^9\msol$)
with respect to previous computations (see Fig.\,\ref{s3f1}).

Following \citet{mer08} we adopt an analytic expression for the local
BHMF, given by the convolution of a Schechter function with a Gaussian
scatter. The Schechter function has the following parametrization:
\beq
\Psi_M=\phi_{\star}\Bigg(\frac{M}{M_{\star}}\Bigg)^{1+\alpha}
\exp\Bigg(1-\frac{M}{M_{\star}}\Bigg)
\label{s3e1}
\eeq
with $\phi_{\star}=10^{-3}$, $\log M_{\star}=8.4,$ and
$\alpha=-1.19$. These parameters are chosen in order to give a mass
function that is coincident with the central value within the
uncertainty range of \citet{sha09} if a Gaussian scatter of
0.3\,dex is used.

In the following analysis, we use a local BHMF computed with a
Gaussian scatter of 0.3 and 0.5\,dex. In this way, we can take into
account the current uncertainties on the local BHMF at the highest
masses. Increasing the scatter from 0.3 to 0.5\,dex, the mass function
becomes compatible with estimates of \citet{sha13b}: the number
density of SMBHs is larger by a factor $\sim2$ at $M\sim10^9\msol$,
and by an order of magnitude at $M\sim6\times10^9\msol$. Changing the
scatter has only, however, a moderate impact on the local SMBH mass density:
$\rho_{BH}=4.3\,(6.6)\,\times10^5\msol\,$Mpc$^{-3}$ for a Gaussian
scatter of 0.3\,(0.5)\,dex.

\subsection{Radiative efficiency}
\label{s3s2}
 
The average radiative efficiency $\epsilon$ of quasars is commonly
estimated based on the classical ``Soltan argument'' \citep{sol82},
according to which the local mass budget of black holes in galactic
nuclei should be accounted for by integrating the overall energy
density released by AGN with an appropriate mass-to-energy conversion
efficiency.  Several studies, assuming a fixed radiative efficiency
over black hole mass and redshift, concluded that $\epsilon$ has to be
within the range 0.05--0.40 to explain the relic population
\citep[e.g.,][]{yu02,elv02,mar04,sha09,li12}. The same range of values
is also predicted by the standard accretion disc theory \citep{sha73},
from $\epsilon=0.054$ for a non--rotating SMBH to $\epsilon=0.42$ for
a SMBH with maximal spin.

Individual AGN can also be used to deduce the absolute accretion rate
and the radiative efficiency by fitting observed optical spectra with
the thin accretion disk model \citep{dav11,wu13}. Average values of
$\epsilon\approx0.1$ are derived, approximately consistent with estimates
based on the Soltan argument.

It is still debated whether or not the radiative efficiency depends on SMBH mass
and redshift. Theoretical arguments have suggested that the radiative
efficiency may increase with the SMBH mass
\citep[e.g.,][]{vol07,fan11}. \citet{dav11} directly determined the
radiative efficiency for 80 quasars and found a strong correlation
with the SMBH mass. On the other hand, \citet{wu13} showed that the
SDSS quasar data are consistent with no intrinsic correlation with the
SMBH mass and no redshift evolution. They also demonstrated that the
apparent correlation $\epsilon$--$M$ can be produced by selection
effects and bias induced by mass estimates \citep[see
also][]{rai12}. Results based on Soltan's argument are also
contrasting: \citet{cao08,wan09,li12} are in favor of a mass and
redshift dependence, while \citet{cao10,sha13} reach opposite
conclusions.

For simplicity, our analysis assumes that $\epsilon$ is constant. We solve the continuity equation for different values
of $\epsilon$, mainly between 0.05 and 0.1 (larger values of
$\epsilon$ give small or negligible evolution of the BHMF in time, and
they are discussed in Appendix\,\ref{app1}).

\subsection{Active SMBHs}
\label{s3s3}

Throughout the paper, we define an active SMBH (or, equivalently, an
AGN) if the Eddington ratio is $\lambda\ge\lambda_{cut}=10^{-4}$. At
lower Eddington ratios, SMBHs are supposed to be quiescent. This choice
is somewhat arbitrary if the Eddington ratio distribution is broad and
extends to $\lambda<10^{-4}$ as indicated by different observations
\citep[e.g.,][]{pan06,hop06,bab07}. We choose this $\lambda_{cut}$
because SMBHs with smaller Eddington ratios do not give relevant
contributions to the bolometric luminosity function, at least in the
luminosity range we consider,
that is, $10^{43}$--$10^{48}$\,erg\,s$^{-1}$. In fact, if
$\lambda<10^{-4}$ , only SMBHs with mass $M\ga10^9\msol$ have luminosity
$L\ge10^{43}$\,erg\,s$^{-1}$. The number density of SMBHs
exponentially declines at these masses, that is, much faster than the
power--law increase of the Eddington ratio distribution at low
$\lambda$ as seen,for example, by \citet{air12}.

The duty cycle is clearly sensitive to the Eddington ratio limit defining
an active SMBH. A lower threshold obviously increases the fraction of
active SMBHs, but, in principle, the BHMF determined by the continuity
equation should be independent of this. However, the evolution of the
BHMF is affected by $\lambda_{cut}$ through the average accretion rate
that is proportional to the product of the duty cycle and $\alamb$
(see Eq.\,\ref{s2e3}). As shown in Appendix\,\ref{app1},
$\lambda_{cut}$ seems to not be particularly relevant for the mass function of
both total and active SMBHs. Moreover, the choice of $\lambda_{cut}$
is still less important when we compare the model predictions with
estimates from observational samples in which quasars are selected
above a given flux limit and that are sensitive only to a specific
range of Eddington ratios.

An alternative view, considered, for example, by \citet{mer08}, assumes that
every SMBH is active at some level, and that there is no distinction
between SMBHs and AGN. By definition, the duty cycle is then equal to
1. However, this implies the knowledge of the Eddington ratio
distribution at very low accretion rates, $\lambda\ll10^{-4}$, which
are scarcely accessible to observations.

\subsection{Eddington ratio distribution}
\label{s3s4}

There are observational evidences that the Eddington ratio
distribution is different for type--1 (or unobscured) and type--2 (or
obscured) AGN. \citet{tru11} proposed two distinct distributions of
Eddington ratios for X-ray-selected AGN, corresponding to two
different modes of accretion. They found that, in general, broad
emission lines are present only at high accretion rates
($\lambda>10^{-2}$). According to this study, unobscured AGN (and possibly
some obscured narrow-line AGN) should be fed by a thin accretion disk
containing the broad-line region and some obscuring material. On the
other hand, narrow--line or lineless AGN at $\lambda<10^{-2}$ may be
powered by a geometrically thick, radiatively inefficient accretion
flow. \citet{lus12} studied the Eddington ratio distribution for an
X--ray sample of type--1 and type--2 AGN covering redshifts
$z\le2$. The Eddington ratio distribution for type--1 AGN was found to
be more consistent with a Gaussian than a power law. For type--2 AGN,
results are less clear, with some evidence for a power--law
distribution only at low redshifts. The presence of two distinct
regimes for the SMBH growth was also pointed out by
\citet{kau09}. They analyzed the observed distribution of Eddington
ratios for a sample of nearby SDSS galaxies. They found that galaxies
with significant star formation are characterized by a broad
log--normal distribution of accretion rates peaked at a few percent of
the Eddington limit. On the contrary, galaxies with old central
stellar populations are characterized by a power--law distribution
function of Eddington ratios.

The Eddington ratio distribution for type--1 AGN was accurately
determined by \citet{she12,kel13} in a large redshift range, from
0.3 to 5. They jointly estimated the BHMF and the Eddington ratio
distribution function (ERDF) for a sample of approximately $58,000$ type--1
quasars from the SDSS. They employed a Bayesian technique to deal with
selection effects and the statistical scatter between true SMBH masses
and virial mass estimates. In particular, \citet{kel13} modeled the
joint bivariate distribution of SMBH mass and Eddington ratio as a
superposition of 2D log--normal functions. Their results on the ERDF
at different redshift bins show that the co-moving number densities of
type--1 quasars increase toward lower Eddington ratios, a trend that
continues beyond the incompleteness limit. The only exceptions are the
$z=3.75$ and $z=4.25$ bins, which display evidence for a peak at
$\lambda\approx0.3$. The distribution of $\lambda$ is relatively
independent of the black hole mass at both low ($z\la0.6$) and high
($z\ga3.2$) redshift, while at intermediate redshifts, they found
larger Eddington ratios moving from $M\sim5\times10^8\msol$ to
$M\sim5\times10^9\msol$. However, as they commented, the mass
dependence could not be real but be driven by systematic effects
related to the change of the mass estimator in these redshift bins.

Concerning type--2 AGN, different independent works agree that the
Eddington ratio distribution has a power law behavior. \citet{hop09}
showed that the observed distributions from the complete sample of
SDSS galaxies selected by \citet{hec04,yu05} are well fitted by a
Schecter function with a power law slope of approximately $-0.6$, almost
independent of the mass range, and a cut--off at
$\lambda\sim1$. \citet{air12} estimated the distribution of the
specific accretion rate, the rate of black hole growth relative to
the stellar mass of the host galaxy, for a large sample of obscured
X--ray AGN at redshift $0.2<z<1.0$. They found that the distribution
is independent of the stellar mass of the host galaxy and can be
described by a power law with constant slope $\simeq-0.65$ throughout
the specific accretion rate ($10^{-4}$--1) and redshift range. These
results can be interpreted in terms of Eddington ratio assuming a
direct proportionality between the mass of the central SMBH and the
stellar mass of the host galaxy (they consider
$M_{BH}\approx0.002M_{\star}$). Similar results were also obtained by
\citet{bon12} from an X--ray and optically selected sample of AGN, but
with a slightly steeper power--law index ($\approx-1$).

Based on these observational findings, we model the
Eddington ratio distribution, $P(\lambda$), for type--1 and type--2
AGN separately. In both cases, we assume distributions independent of the SMBH
mass. According to our definition of active SMBHs, $P(\lambda)$ is only
considered for $\lambda\ge10^{-4}$.

\begin{figure}
  \centering
  \includegraphics[width=\hsize]{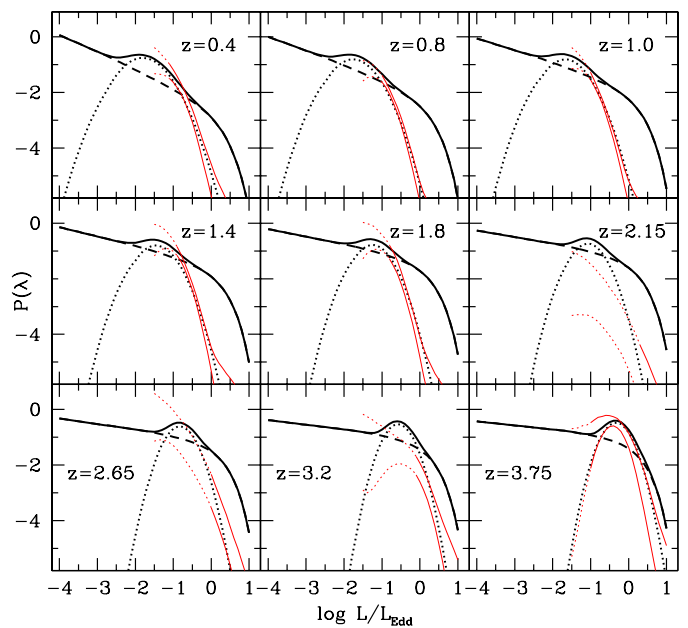}
  \caption{Eddington ratio distribution given by Eq\,.\ref{s3e4} for
    SMBHs of mass $M=10^8\msol$ at different redshifts (black solid
    lines; dotted lines are for type--1 AGN and dashed
    lines for type--2 AGN). The Eddington ratio distributions from
    \citet{kel13} with an arbritary normalization are also shown (red
    lines): the two lines give the uncertainty in their estimates (at
    68\% of probability); dotted lines denote the regions below the
    10\% completeness for the flux--limited SDSS sample.}
  \label{s3f3}
\end{figure}

For type--1 AGN, we use a log-normal distribution,
\beq
P_1(\lambda,z)=\frac{1}{2\pi\sigma(z)\lambda}\,
e^{-[\ln\lambda-\ln\lambda_c(z)]^2/2\sigma^2(z)}\,,
\label{s3e2}
\eeq
where the central value $\lambda_c(z)$ and the dispersion $\sigma(z)$
of the distribution are determined by fitting the shape of the ERDFs from
\citet{kel13} in the different redshift bins and by interpolating the
results with a linear function (see Fig.\,\ref{s3f3}). We find
\beq
\begin{array}{ccl}
\log\lambda_c(z) & = & \max[-1.9+0.45z,\log(0.03)] \nonumber \\
\sigma(z) & = & \max(1.03-0.15z,0.6)\,.
\label{s3e2a}
\end{array}
\eeq

\begin{figure}
  \centering
  \includegraphics[width=\hsize]{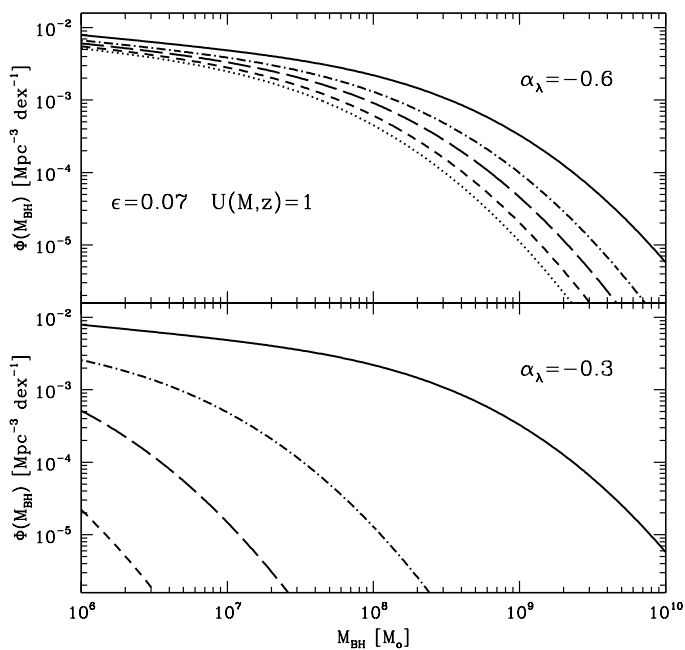}
  \caption{SMBH mass function at redshifts 0, 0.5, 1, 2, and 4 (from
    top to bottom), assuming $U(M,z)=1$ and a power--law distribution
    for the Eddington ratio with a constant slope
    $\alpha_{\lambda}=-0.6$ (upper panel) and $-0.3$ (lower panel).}
  \label{s3f3b}
\end{figure}

For type--2 AGN, we use a power--law distribution with an exponential
cut--off at super--Eddington luminosities,
\beq
P_2(\lambda,z)=a_2(z)\,\lambda^{\alpha_{\lambda}(z)}
e^{-\lambda/\lambda_0}\,,
\label{s3e3}
\eeq
where $\lambda_0=1.5$ (or 2.5 when $\epsilon\ge0.1$). The
distribution is normalized to unity by the factor $a_2$. The slope of
the power law is taken to be
\beq
\alpha_{\lambda}=\left\{ \begin{array}{ll}
    -0.6 & z\le0.6 \\
    -0.6/(0.4+z) & z>0.6
  \end{array} \right.
\label{s3e3b}
.\eeq
At low redshifts, this is in agreement with findings from
\citet{hop09,kau09,air12}. At redshifts $z\ga1,$ there are no
observational constraints on the Eddington ratio distribution for
type--2 AGN. Here we introduce a redshift dependence in
$\alpha_{\lambda}$ that makes the distribution flatter and flatter at
high redshifts. As discussed in Appendix\,\ref{app1}, keeping a
constant slope of $-0.6$ would give a BHMF that does not significantly
evolve over time at $z\ga0.5 $ and at low and intermediate masses, in
contrast with general expectations. For example, Fig.\,\ref{s3f3b}
shows the evolution of the BHMF computed from the continuity equation,
assuming $U(M,t)=1$ and a power--law Eddington ratio distribution with
constant slope $\alpha_{\lambda}=-0.6$ and $-0.3$. Even with a duty
cycle equal to 1, most of SMBHs with $M<10^8\msol$ should already be
formed before $z=4$ if $\alpha_{\lambda}=-0.6$. To be noted, the large
difference in the mass function evolution between the two values of
$\alpha_{\lambda}$.

As explained in \citet{hop09}, given the quasar luminosity function,
the Eddington ratio distribution can be directly translated into a
quasar lifetime or light curve model. A log-normal distribution, as
we use for type--1 AGN, is typically associated to ``light bulb''
models, in which quasars grow at fixed Eddington ratio with an
instantaneous or exponential luminosity decay. A truncated power-law
Eddington distribution arises instead if the quasar luminosity undergoes
a power--law decay with time. Our assumptions on $P(\lambda)$
imply therefore a different accretion model for type--1 and type--2
AGN.

The Eddington ratio distribution for all the active AGN is thus the
sum of the distributions for type--1 and type--2 quasars weighted by
the relative abundance of the two populations. The fraction of
obscured/unobscured AGN has been largely investigated in the literature,
providing evidence of an anti-correlation with nuclear luminosity
\citep[e.g.,][]{ued03,has08,mer14,ued14}. Several works have also
reported evidence of positive evolution of the fraction of obscured
AGN with increasing redshift \citep{laf05,has08,iwa12,mer14}. In the
following analysis, we adopt the parametrization provided by
\citet{ued14} for the fraction of obscured AGN, that is, for AGN with an
intrinsic absorption $N_H\ge10^{22}$\,cm$^{-2}$. This is based on a
combined sample of X--ray surveys of various depths, widths and energy
bands. They provided the fraction of obscured AGN, $f_{obs}$, as a
function of X--ray luminosity and redshift. 

Because the relative abundance of type--1 and type--2 AGN depends on
the luminosity, the total Eddington ratio distribution will depend on
the SMBH mass too and will take the form:
\beq
P(\lambda|M,z) = a_n(M,z)\,\bigg[f_{uno}(L,z)P_1(\lambda,z)+
f_{obs}(L,z)\,P_2(\lambda,z)\bigg]\,,
\label{s3e4}
\eeq
where $f_{uno}(L,z)=1-f_{obs}(L,z)$ is the fraction of type--1 AGN and
$f_{obs}$ is in terms of the bolometric luminosity. X--ray
luminosities are converted to bolometric luminosities through the
bolometric corrections provided by \citet{hop07}. The factor $a_n(M)$
is required by the normalization condition $\int d\log
\lambda\,P(\lambda|M,z)=1$.

Fig.\,\ref{s3f3} shows $P(\lambda)$ for $M=10^8\msol$ at different
redshifts. At low redshift, most SMBHs, including type--1 AGN, accrete
at low rates, $\lambda\ll0.1$. Increasing the redshift, the
distribution for type--2 AGN becomes flatter, while for type--1 AGN
this distribution becomes sharper and peaked at higher $\lambda$. At
$z\ga2,$ almost all the type--1 AGN and approximately 30--50\%
of the total active population have $\lambda>0.01$. At this redshift,
the fraction of AGN with $L\sim L_{Edd}$ also becomes relevant. The
dependence of the distribution on the SMBH mass is modest, and is
related only to the increasing fraction of type--1 AGN with $M$.

\begin{figure}
  \centering
  \includegraphics[width=\hsize]{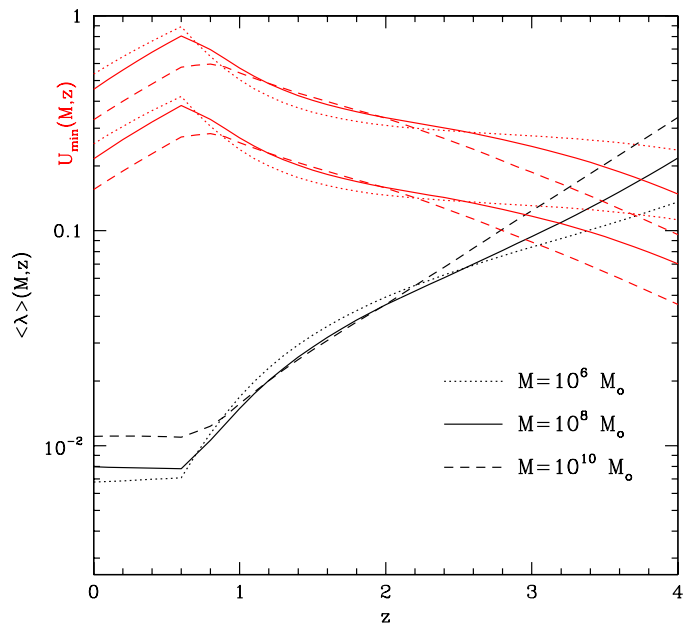}
  \caption{Redshift evolution of the average Eddington ratio (black
    lines), and of the duty cycle required in order to have a growth
    time equal to the age of the Universe (red curves). The duty cycle
    is computed for $\epsilon=0.1$ (upper lines) and 0.05 (lower
    lines). We have considered three SMBH masses: $10^6$ (dotted
    line), $10^8$ (solid line), and $10^{10}\msol$ (dashed line).}
  \label{s3f4}
\end{figure}

Given the probability distribution, it is easy to compute the average
Eddington ratio as a function of black hole mass and redshift
(Fig.\,\ref{s3f4}). We find that it is approximately $10^{-2}$ at $z<1$ and
then steadily increases with redshift up to 0.1--0.3 between $z=$3 and $z=$4.
As expected, there is only a slight dependence on the SMBH mass.

\subsection{Quasar luminosity function}
\label{s3s5}

\begin{figure}
  \centering
  \includegraphics[width=\hsize]{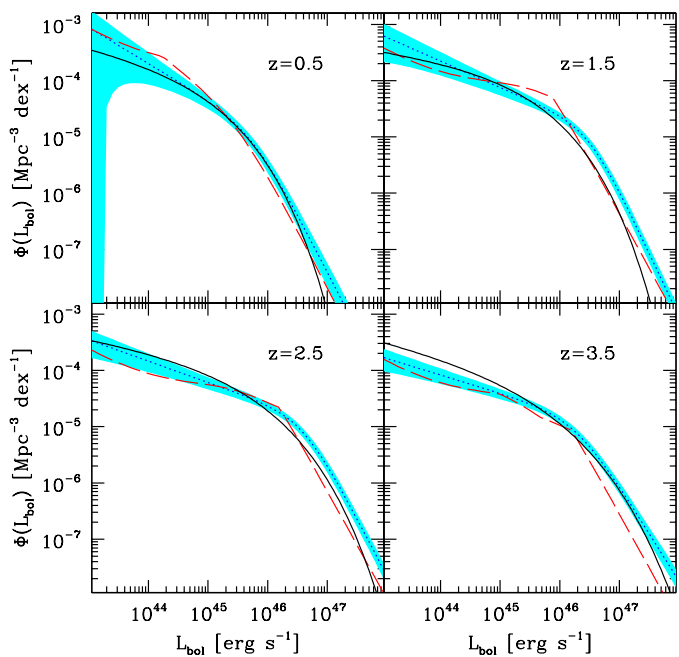}
  \caption{Bolometric luminosity function of quasars at different
    redshifts from: \citet[][blue dotted lines plus cyan shaded
    areas for the 1--$\sigma$ uncertainty]{hop07}; \citet[][red
    dashed lines]{ued14}; and by \citet[][black solid lines]{ave15}.}
  \label{s3f5}
\end{figure}

We adopt the fitting function to the bolometric quasar luminosity
function (QLF) provided by \citet{hop07}. They combined a large set of
measurements from the optical, soft, and hard X--ray, and near-- and
mid--IR bands to determine the bolometric QLF in the redshift interval
$z=0$--6. Then, they fit the observational QLF by a double power law
in each redshift bin. The evolution in redshift of the best-fit
parameters is also parametrized in order to provide analytical
formulas to compute the QLF in a generic redshift between 0 and
6. However, because of the increasing uncertainties in the QLF at high
redshifts, we decided to restrict our analysis up to a redshift
of 4 only. We use the results for the reference model, indicated as ``full''
in Table\,3 of \citet{hop07}. Based on the uncertainties of the
best--fit parameters, we estimate the uncertainty on the QLF by a
Montecarlo technique (see the shaded areas in Fig.\,\ref{s3f5} and
\ref{s4f1}).

It is informative to compare the Hopkins et al. QLF, based on data
available up to 2007, to QLFs derived from more recent data. In
Figure\,\ref{s3f5}, we plot the bolometric luminosity functions
obtained by (1) \citet{ued14}, based on surveys in the soft and hard
X--ray bands; and (2) \citet{ave15}, built up from a compilation of
observations in the optical and X--ray bands. We find good
consistency, although the Hopkins et al. QLF is slightly higher than
the other estimates at high luminosities
($L\ga10^{46}$\,erg\,s$^{-1}$).

\begin{figure}
  \centering
  \includegraphics[width=\hsize]{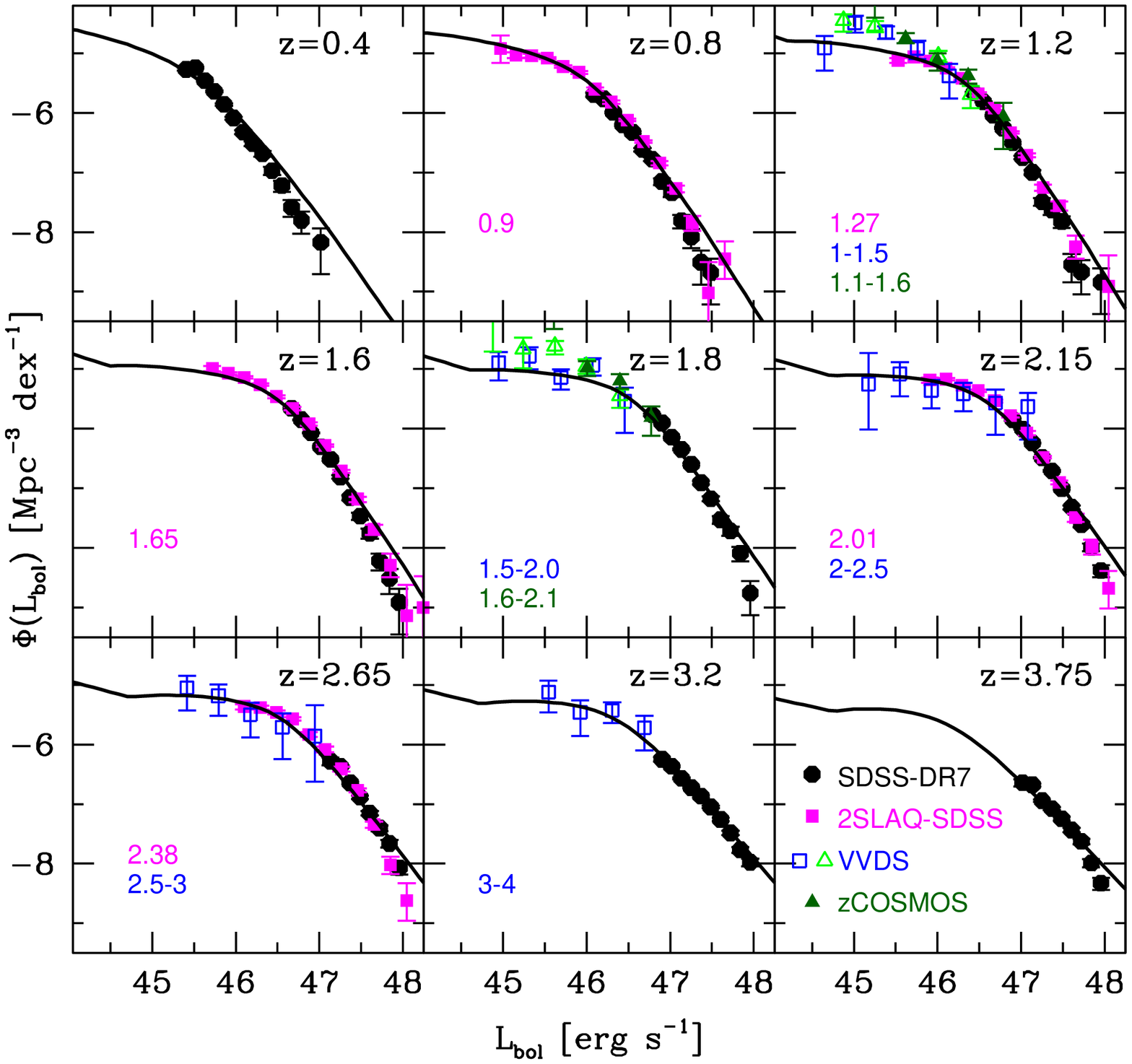}
  \caption{Bolometric luminosity function for type--1 AGN (black
    lines) obtained from the QLF of \citet{hop07} rescaled by the
    fraction of unobscured objects of \citet{ued14}. The points
    represent measurements from different optical surveys: SDSS DR7
    \citep[black,][]{she12}; 2SLAQ--SDSS \citep[magenta,][]{cro09};
    VVDS (blue, \citealt{bon07} and green, \citealt{sch15}); zCOSMOS
    \citep[dark green,][]{sch15}. The redshift for the model and for
    the SDSS DR7 data is marked in the upper right corner of each
    plot, and the redshifts for other data are marked in the lower
    left corner in the corresponding colors when they are different.}
  \label{s3f6}
\end{figure}

As an additional check, in Figure\,\ref{s3f6}, we plot the luminosity
function of type--1 AGN obtained from the Hopkins et al. QLF
multiplied by the fraction of unobscured AGN used in our analysis
\citep[i.e., from][]{ued14}. This is compared to observational
estimates of type--1 AGN luminosity functions from different optical
surveys \citep{bon07,cro09,she12,sch15}\footnote{Optical magnitudes of
  the different data are converted before to 2500\,$\AA$ continuum
  luminosities using the relations provided by Eq.\,19 in
  \citet{she12} and then to B--band luminosities assuming a power--law
  continuum slope $\alpha_{\nu}\simeq-0.5$. The final conversion to
  bolometric luminosities is done through the bolometric corrections
  of \citet{hop07}.}. The agreement is, in general, extremely
good. This is particularly remarkable considering the different types
of data used to estimate the absorption function (X--ray band) and the
type--1 AGN lumiosity function (optical band).

\subsection{Duty cycle}
\label{s3s6}

As previously discussed, we define the duty cycle $U(M,z)$ as the
fraction of SMBHs of mass $M$ that are active at redshift $z$, or, in
other terms, as the fraction of SMBHs accreting at
$\lambda\ge10^{-4}$. Alternatively, the duty cycle of a SMBH can
also be viewed as the lifetime that quasars of a given mass pass
radiating at a luminosity larger than a certain value $L$
\citep[e.g.,][]{hop09}. This is a key term to solving the continuity
equation and is also an important prediction of the model.

\citet{mer08} defined the average {growth time} of a SMBH as the
ratio $M/\arate$. It measures the time it would take a SMBH of mass
$M$ to double its mass if accreting at a rate $\arate$. In
Fig.\,\ref{s3f4}, we plot the value of the duty cycle needed to have a
growth time equal to the age of the Universe at that redshift. This
indicates the `minimum' duty cycle above which SMBHs (of mass $M$ at
time $t$) are, on average, actively growing or, in other terms, the
BHMF is significantly evolving. We can see that phases of major
growth/evolution require duty cycles close to 1 at low redshifts and,
in any case, larger than 0.1.

From an observational point of view, the duty cycle of SMBHs is
difficult to estimate because it requires the simultaneous knowledge of
the SMBH and AGN mass function. If we assume that all massive galaxies
contain a SMBH, then the fraction of active galaxies can be used as an
observable proxy for black hole duty cycles. However, even in this
way, the `measured' duty cycle will depend on the luminosity or
Eddington ratio threshold of the specific observational sample.

Our approach is to assume a parametric function for the duty cycle,
which is described by a double power law:
\beq
U(M,z)=\min\Bigg(\frac{A(z)}{(M/M_0(z))^{\alpha_l(z)}+
(M/M_0(z))^{\alpha_k(z)}}\,,1\Bigg)\,,
\label{s3e6}
\eeq
where we have imposed the physical condition that $U(M,z)\le1$. 
The redshift evolution of the duty cycle parameters is described by
cubic polynomials,
\beq
X(z)=a_X+b_X\,z+c_X\,z^2+d_X\,z^3~~~~{\rm with}~
X=A,\,\alpha_l,\,\alpha_k,\,M_0,
\label{s3e7}
\eeq 
giving a total of 16 parameters for the duty cycle. These
parameters are constrained by the condition that, at each redshift
bin, the bolometric luminosity function computed by Eq.\,\ref{s2e7}
%
%
has to match the observational QLF of \citet{hop07}.

More in detail, given the duty cycle (i.e., chosen a set of the 16
parameters), the continuity equation is solved recursively, imposing
the local SMBH mass function as an initial condition and going backwards
in time from redshift 0 to 4. At each timestep, we use the AGN MF
derived from the previous timesteps to compute the update BHMF. The
bolometric QLF is finally computed from Eq.\,\ref{s2e7}. We employ a
Markov Chain Monte Carlo (MCMC) method to find the parameters of the
duty cycle that best fit the QLF of \citet{hop07} in the redshift
range 0--4.

In principle, if the Eddington ratio distribution is fixed, the AGN MF
could be directly derived from the convolution of the input
$P(\lambda)$ and the QLF. The continuity equation would be used only
to determine the BHMF evolution. However, this procedure does not
guarantee the condition for the duty cycle to be $U(M,z)\le1$. We have
verified, in fact, that this is not the case when the AGN MF is
parametrized by a double power law \citep[that is the most natural
choice as the QLF is also well described by a double power law; see,
e.g.,][]{cao10,sha13}.

As a consistency test, we developed another method to determine
the duty cycle parameters of Eq.\,\ref{s3e6} that does not rely on any
functional form for the redshift dependence of the parameters. In this
case, the duty cycle is assumed to be constant in small redshift
intervals of $\Delta z=0.1$. In each redshift bin, $z_i$, the four free
parameters of $U(M,t)$ (i.e., $A$, $\alpha_l$, $\alpha_k$ and $M_0$)
are determined by finding the best fit to the observational QLF at
$z=z_{i-1}+\Delta z/2$. In general, the two methods provide consistent
results, although the former is able to provide better fits to the
QLF at high redshifts, especially when the BHMF is strongly evolving
with time (i.e., for low values of the radiative efficiency).

\section{Results}
\label{sec4}

In this section, we present the results of the model in terms of BHMF,
duty cycle, and AGN MF (Fig.\,\ref{s4f2}). The average radiative
efficiency of SMBHs is a free parameter and we solved the
continuity equation for different values of $\epsilon$. In addition,
we considered two different inputs of the local mass function,
obtained by convolving the Schechter function in
Eq.\,\ref{s3e1} with a Gaussian scatter of 0.3 (LMF03) and 0.5\,dex
(LMF05).

Three models are described in detail: the LMF05 local BHMF and
$\epsilon=0.05$ and 0.07; the LMF03 local BHMF and
$\epsilon=0.1$. They provide the best fit to the QLF and show more
`reliable' behaviors in term of black hole MF evolution with respect
to other model inputs (see Appendix\,\ref{app1}). Below, we refer
to these models just for the different value of the radiative
efficiency, but encourage the reader to also keep in mind the different
local BHMF used. In Appendix\,\ref{app1}, we show results with
different combinations of $\epsilon$ and the local BHMF, and we
provide a general discussion about the uncertainties in the model
predictions.

\begin{figure*}
  \centering
  \includegraphics[width=6cm]{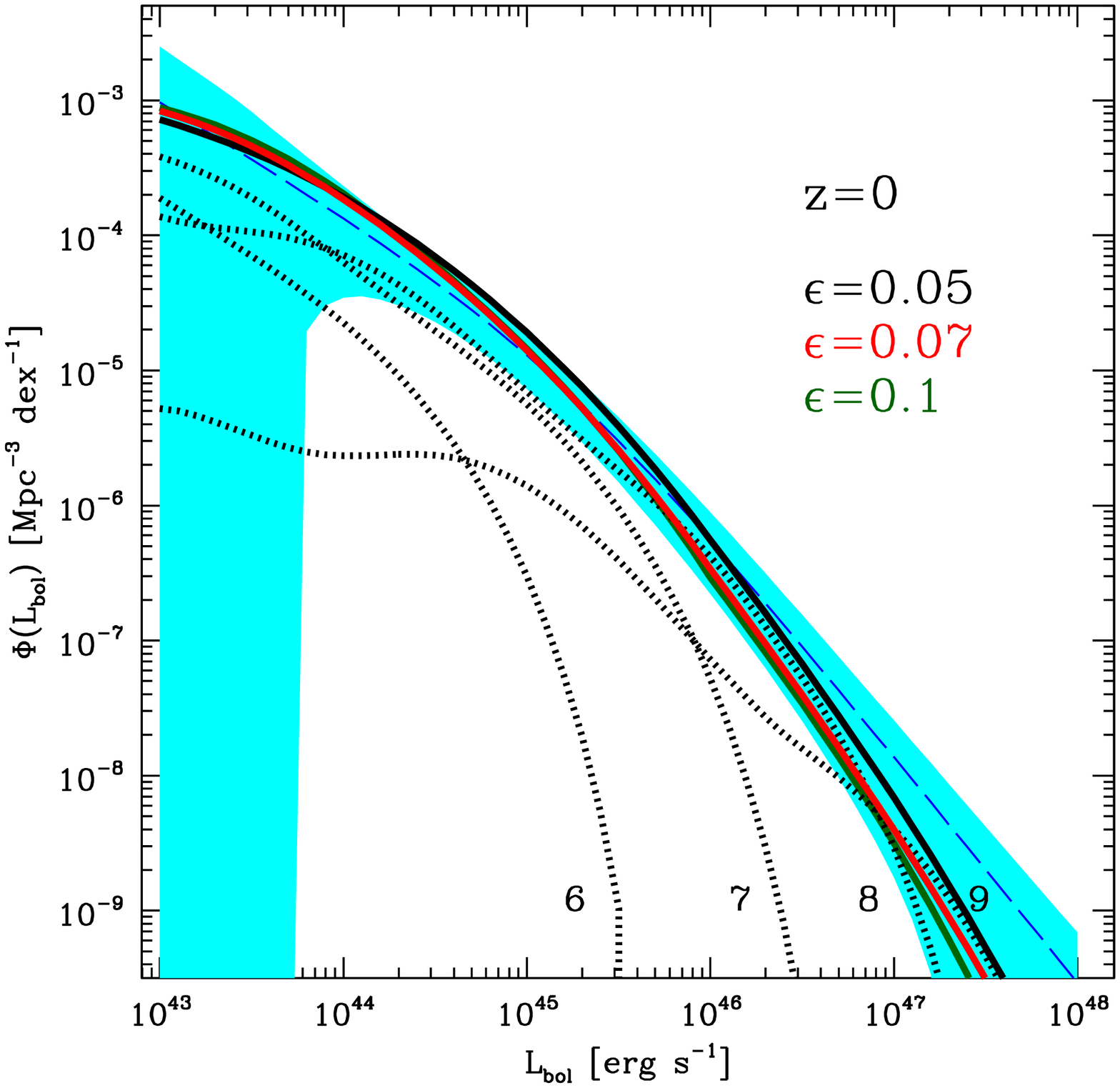}
  \includegraphics[width=6cm]{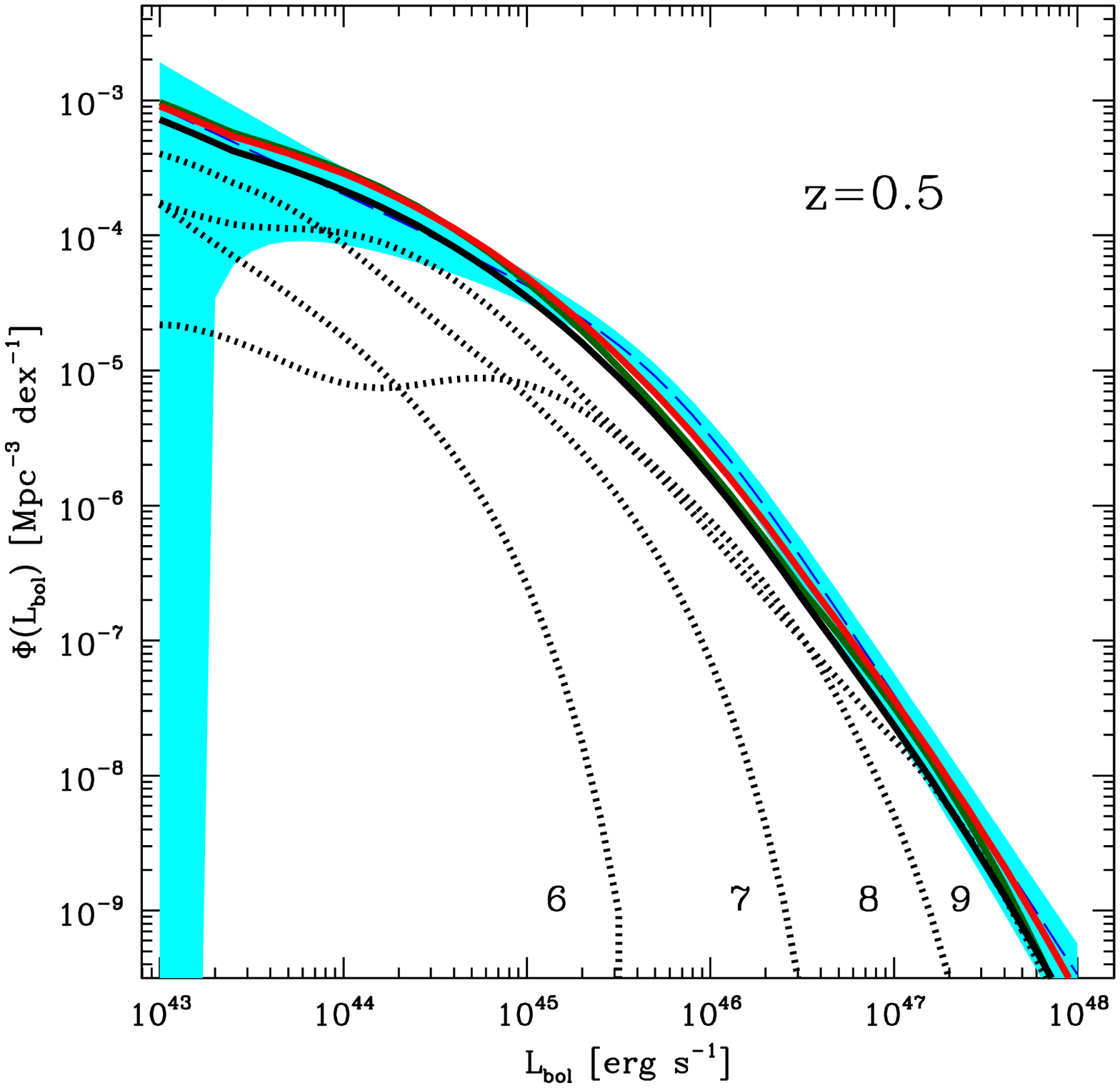}
  \includegraphics[width=6cm]{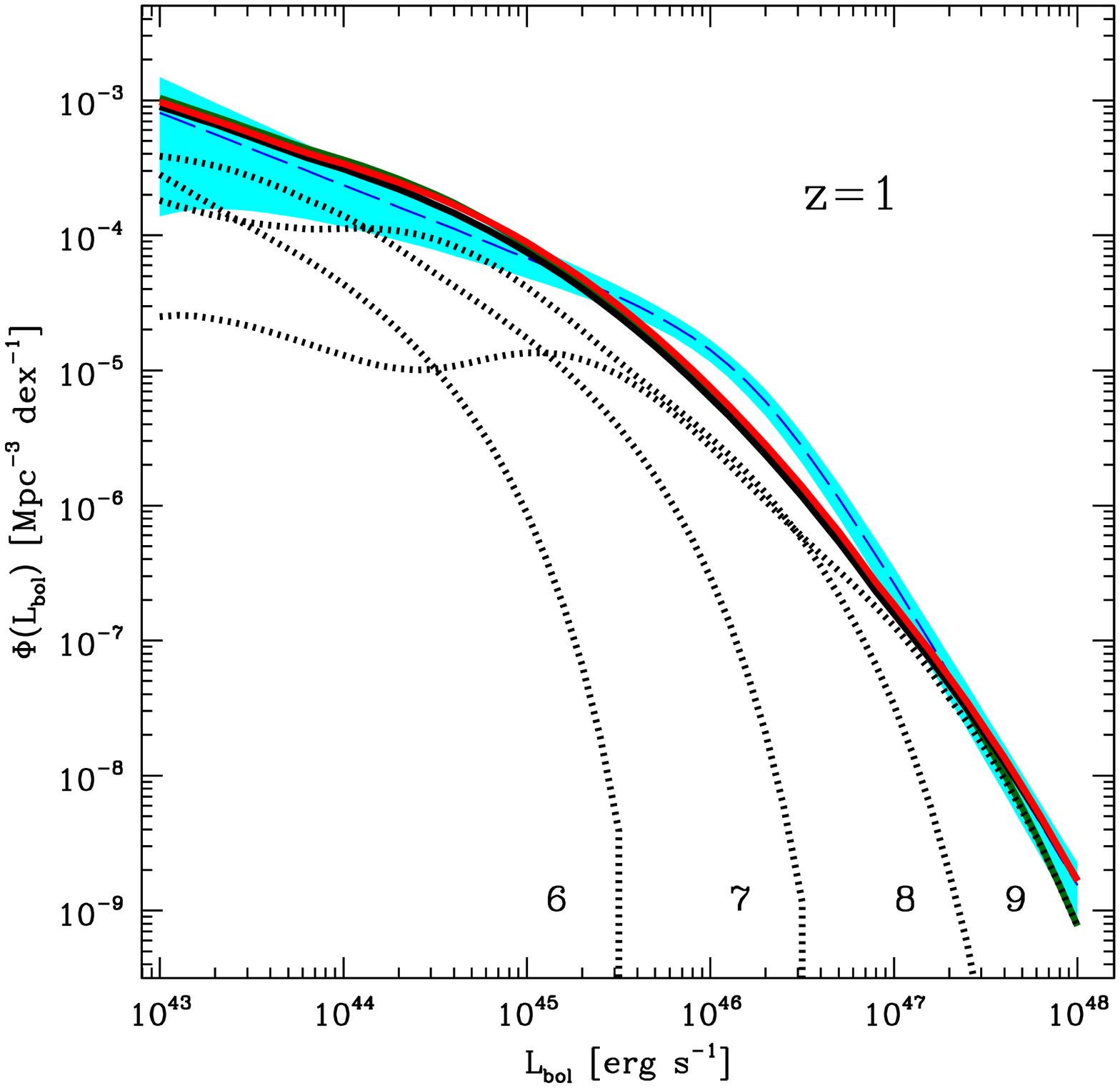}
  \includegraphics[width=6cm]{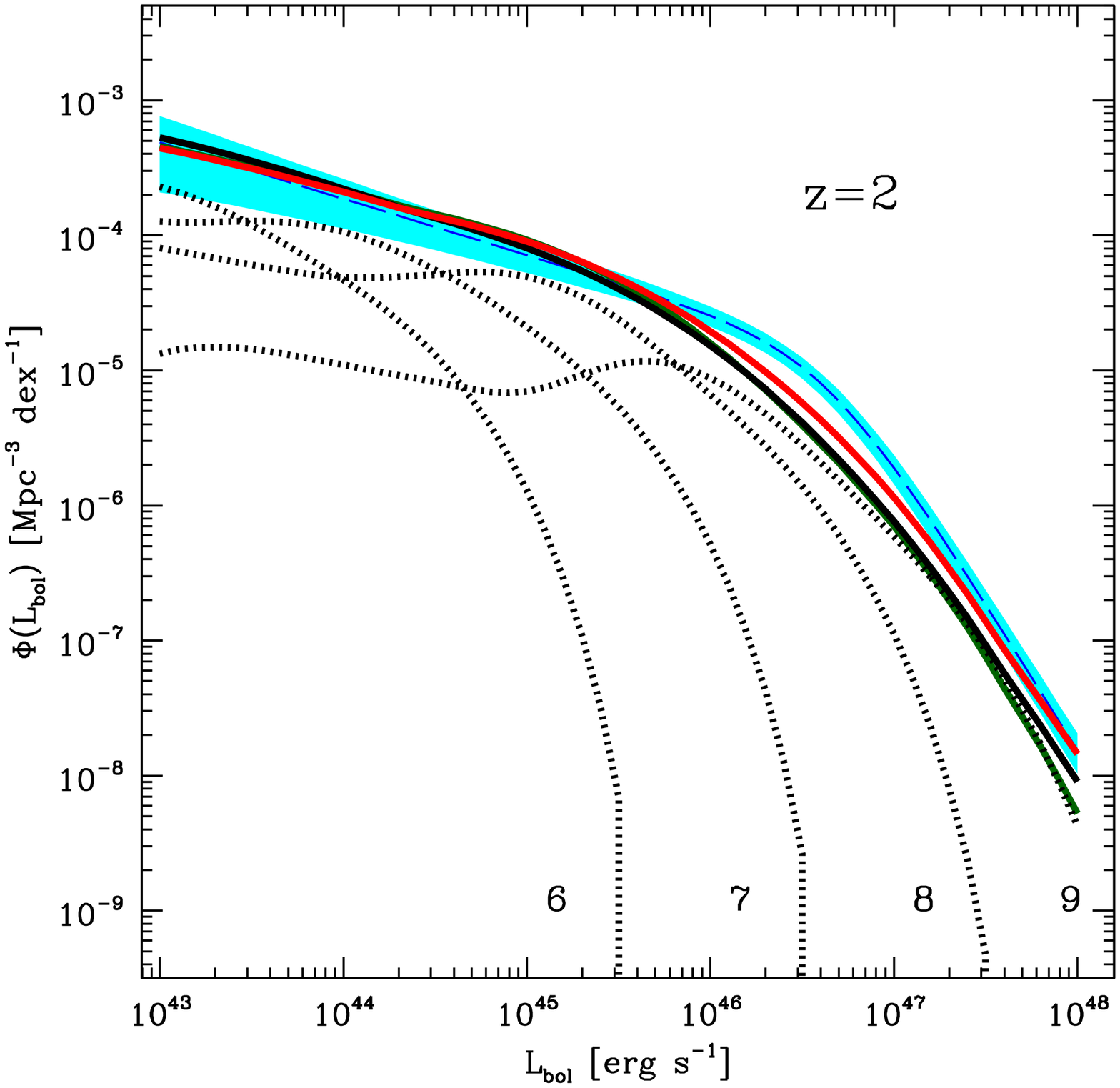}
  \includegraphics[width=6cm]{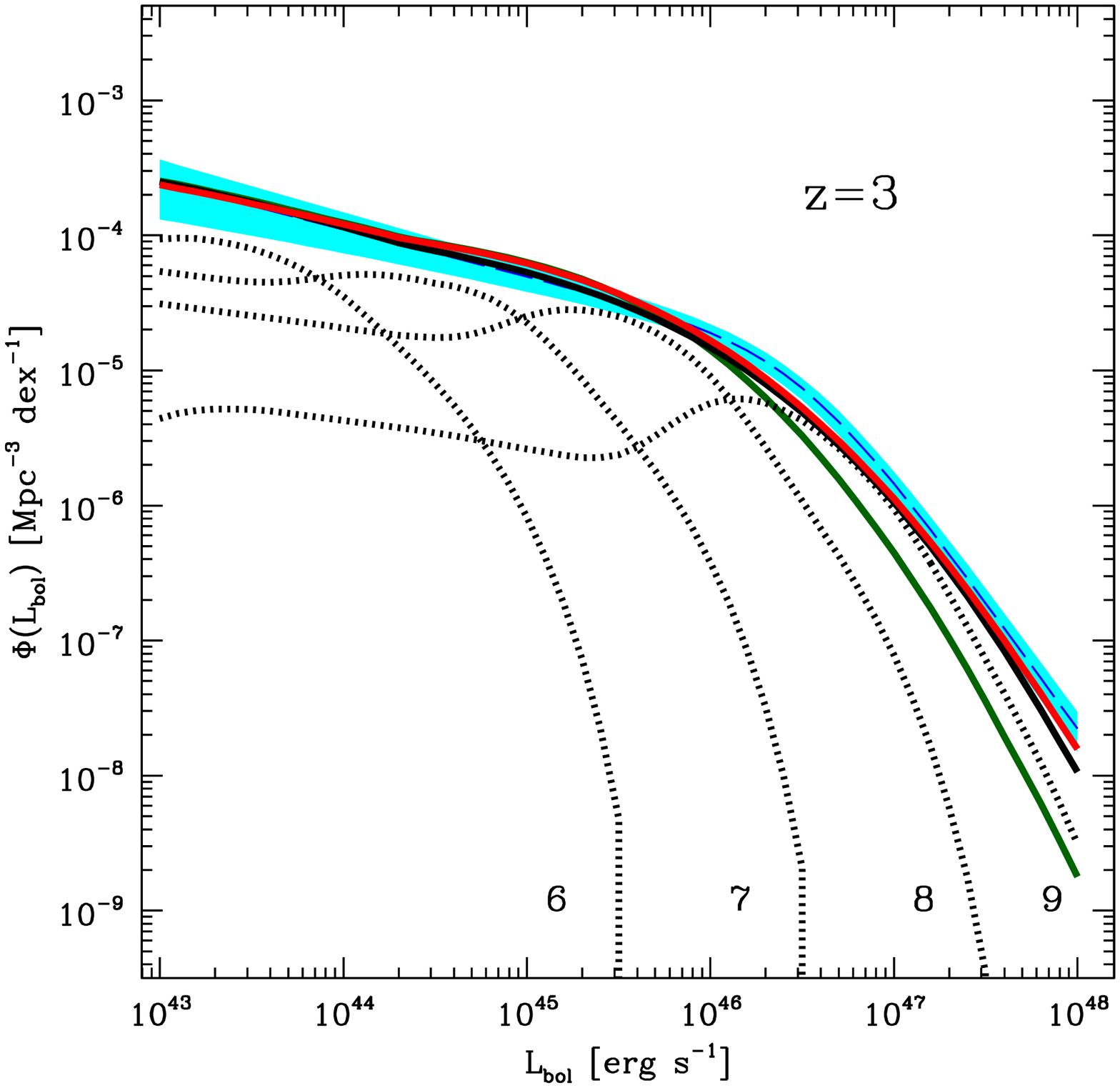}
  \includegraphics[width=6cm]{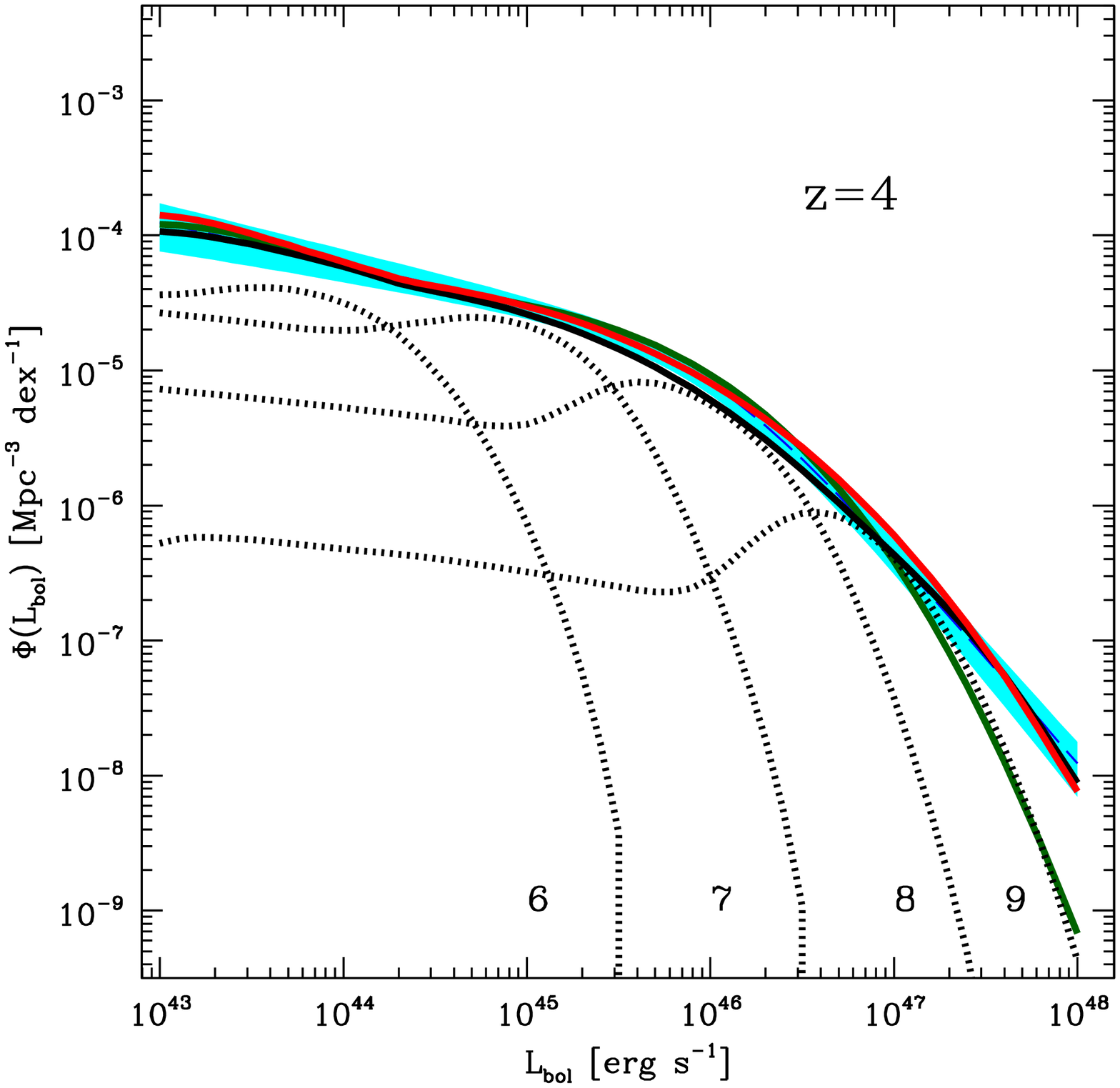}
  \caption{Bolometric quasar luminosity function at different
    redshifts predicted by the models (solid black, red, and green
    lines for $\epsilon=0.05,$ 0.07, and 0.1, respectively), and
    compared with the estimates from \citet{hop07} (blue dashed lines
    plus cyan shaded areas for the 1--$\sigma$ uncertainty). Black
    dotted lines show the contribution to the LF from SMBHs with mass
    in the interval indicated in the plots (e.g., ``6'' corresponds to
    a mass interval $[5\times10^5,\,5\times10^6)\msol$) for the model
    with $\epsilon=0.05$.}
  \label{s4f1}
\end{figure*}

\begin{figure}
  \centering
  \includegraphics[width=\hsize]{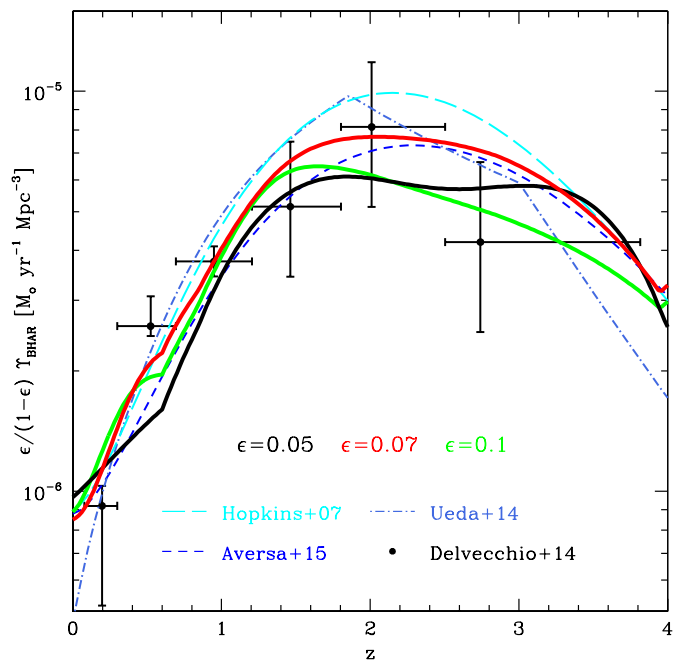}
  \caption{Black hole accretion rate density (multiplied by
    $\epsilon/(1-\epsilon)$) from our reference models with
    $\epsilon=0.05,$ 0.07, and 0.1 (solid thick lines); these are
    compared with the BHAD obtained from different bolometric LFs (as
    indicated in the plot).}
  \label{s4f1b}
\end{figure}

\subsection{Bolometric quasar luminosity function}
\label{s4s1}

In Fig.\,\ref{s4f1}, we compare the QLF estimated by \citet{hop07}
with results from the three models. In general, the models
accurately reproduce the observational LF, although the transition
between the low- and high-luminosity tails is smoother in the models
at $1\la z\la2$. At redshifts $z>2$, the case with $\epsilon=0.1$
tends to underestimate the observational QLF for
$L\ga10^{47}\,$erg\,s$^{-1}$, due to the very low number density of
SMBHs with $M>10^9\msol$ predicted by the model (see below).

In Fig.\,\ref{s4f1}, we also plot the contribution to the QLF from
SMBHs in different ranges of mass. The shape of these contributions
reflects the Eddington ratio distribution. For a small interval of
masses, in fact, the QLF is simply $\Phi(L)\approx
P(\lambda|M)\Phi_{AGN}(M)$.  The peak of the
log--normal distribution of type--1 AGNs is clearly visible and, at high redshifts, we
can associate a narrow range of masses to each luminosity for the
SMBHs that mainly contribute to the QLF.

Finally, we derive the time evolution of the black hole accretion rate
density (BHAD), $\Upsilon_{BHAD}$, as predicted by the models. This
quantity is defined as
\beq
\Upsilon_{BHAD}(z)=\int_0^{\infty}\frac{1-\epsilon}{\epsilon
  c^2}\,L\Phi(L,z)d\log L\,.
\label{s4e0}
\eeq
In Fig.\,\ref{s4f1b}, we plot $\Upsilon_{BHAD}$, normalized by the
factor $\epsilon/(1-\epsilon)$ in order to remove the dependence on
the radiative efficiency. For comparison, we plot the BHAD obtained by
integrating the observational QLFs of \citet{hop07,ued14,ave15}. We
also report the estimates of \citet{del14}, using a sample of
far--infrared galaxies from {\it Herschel}. Our results are highly
consistent with these estimates. At redshift $z\sim2$, where the peak
of the accretion rate density is found, our models underestimate the
BHAD of \citet{hop07} by 20--40\%, but are highly compatible with
estimates of \citet{del14,ave15}.

\subsection{Mass function and duty cycle of SMBHs}
\label{s4s2}

\begin{figure*}
  \centering
  \includegraphics[width=6cm]{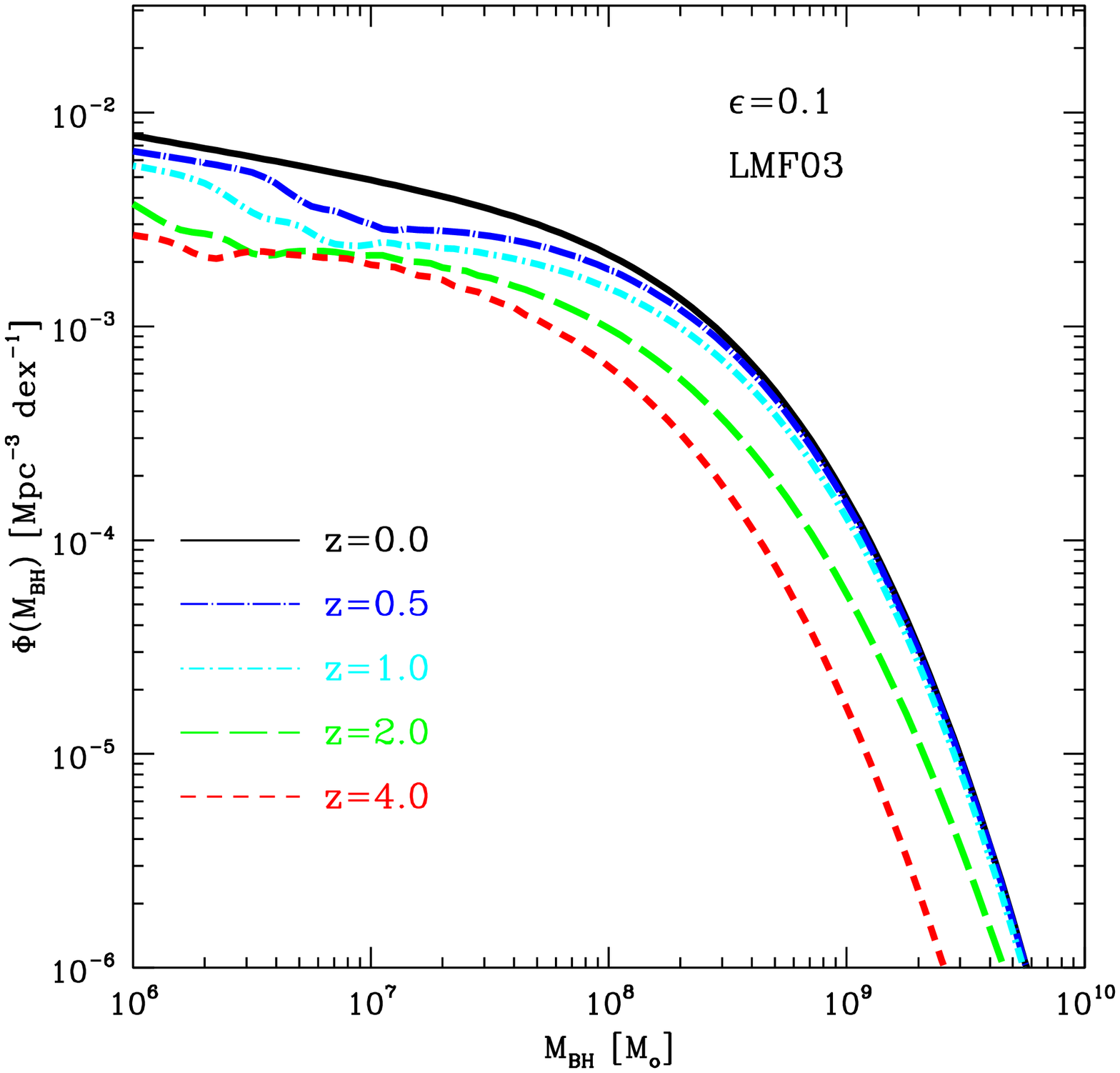}
  \includegraphics[width=6cm]{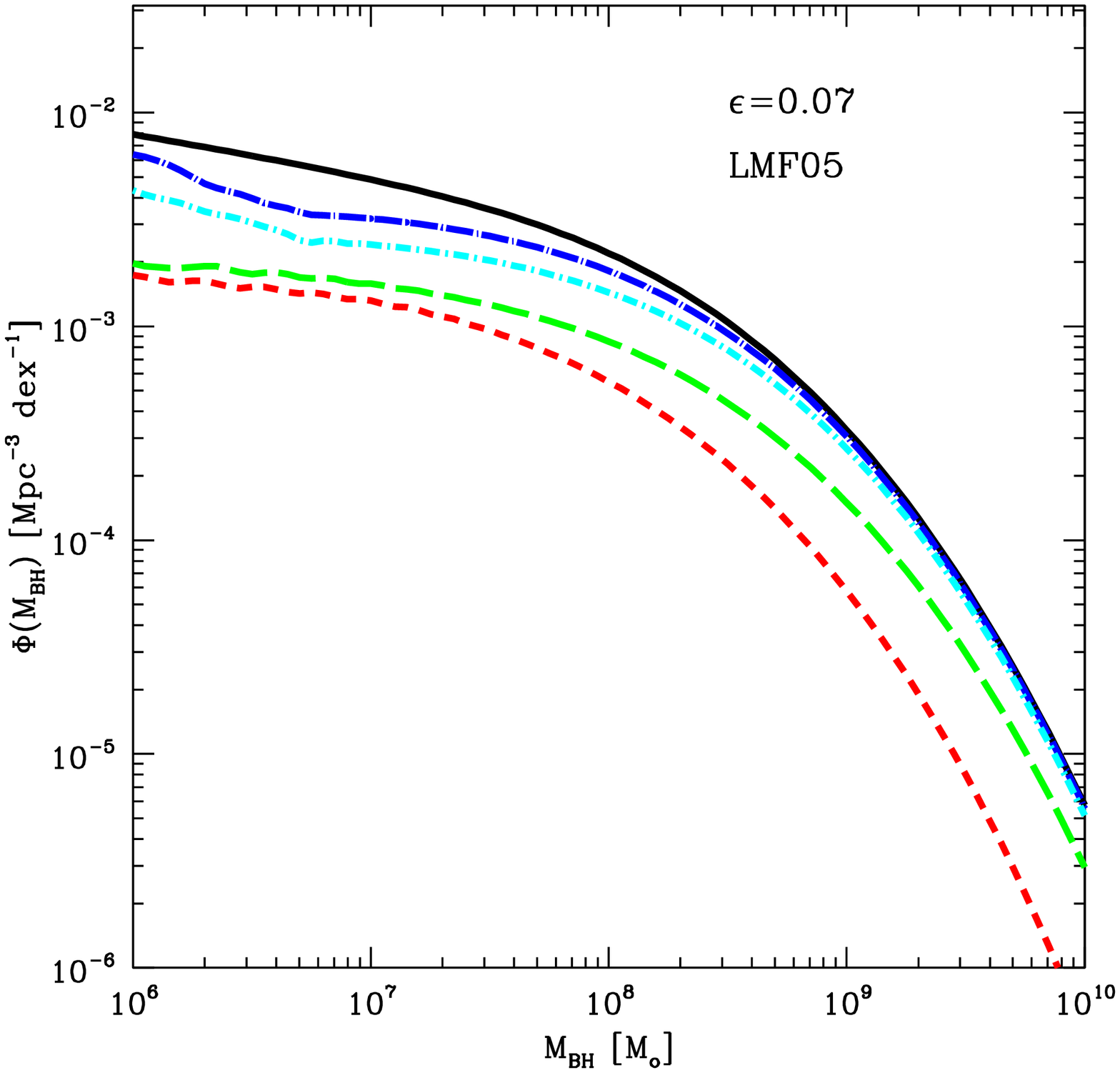}
  \includegraphics[width=6cm]{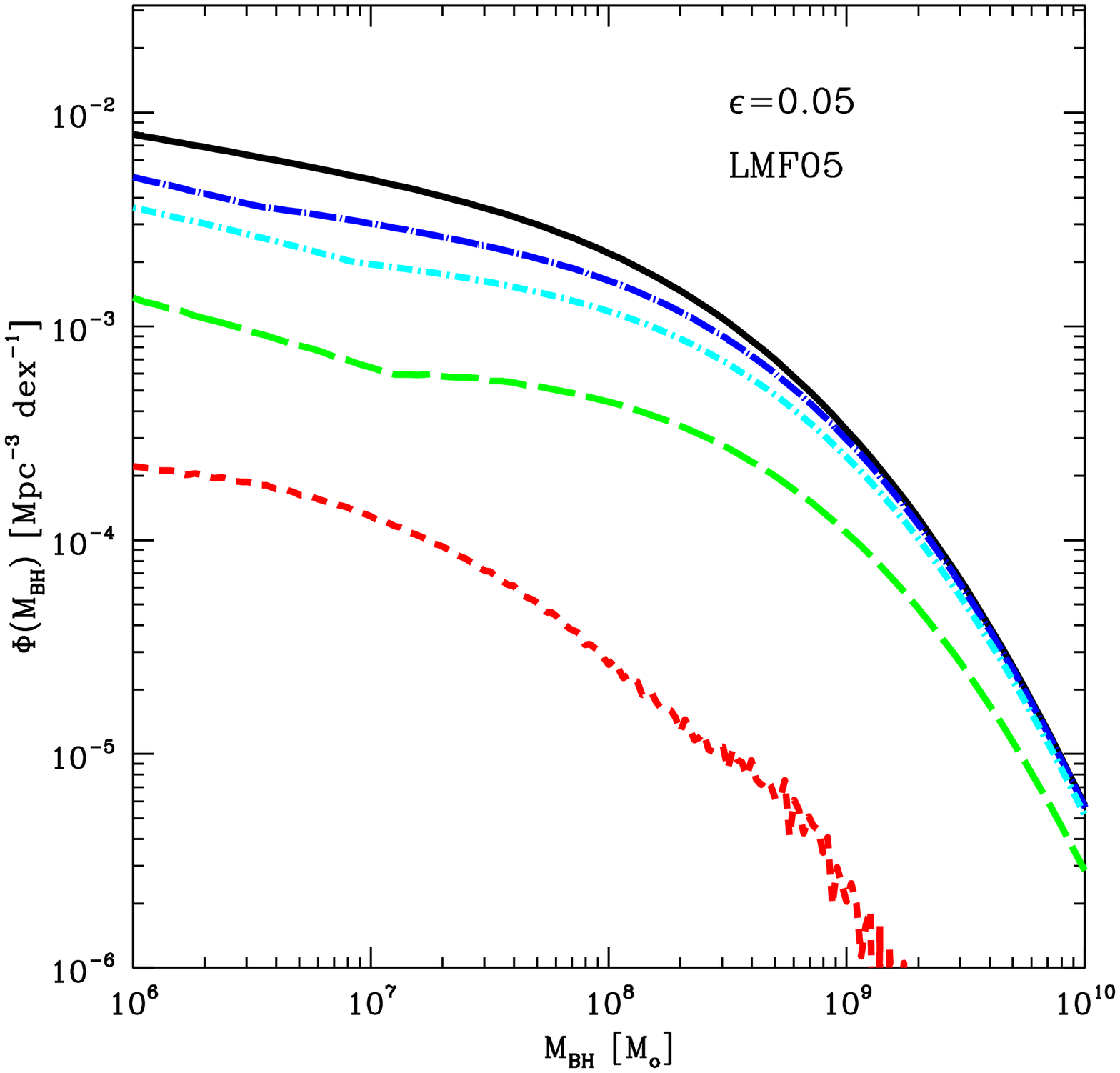}
  \includegraphics[width=6cm]{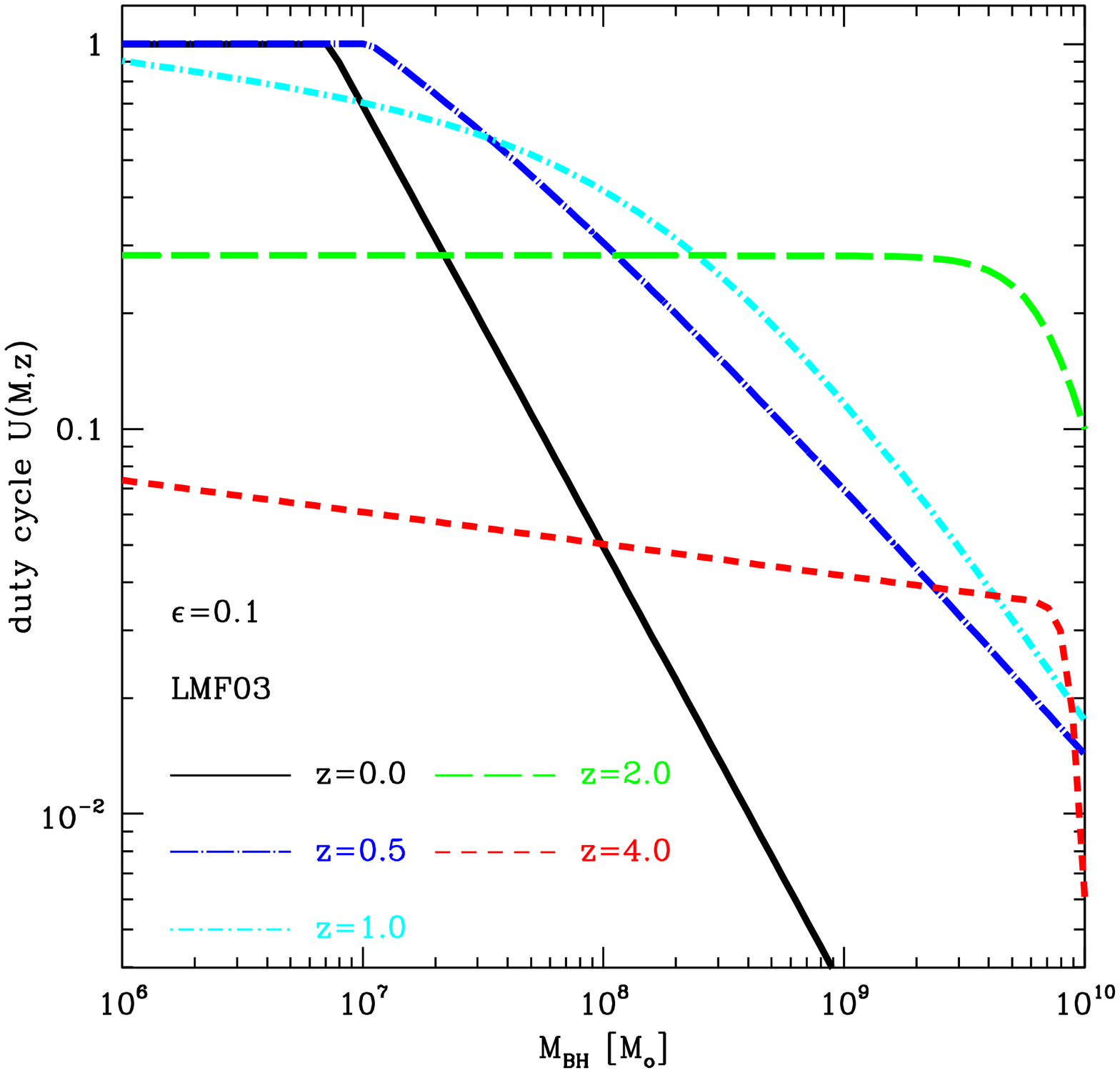}
  \includegraphics[width=6cm]{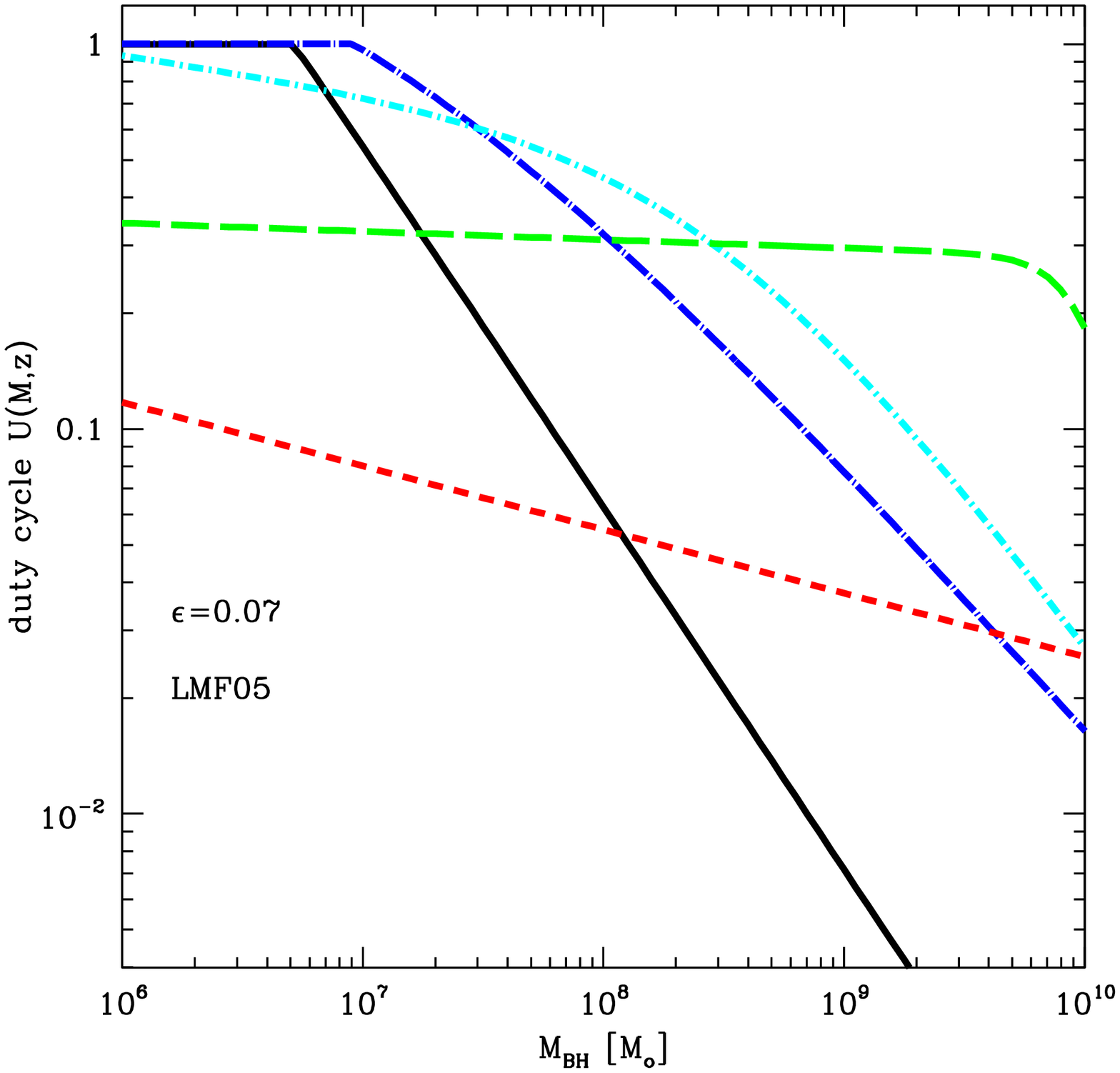}
  \includegraphics[width=6cm]{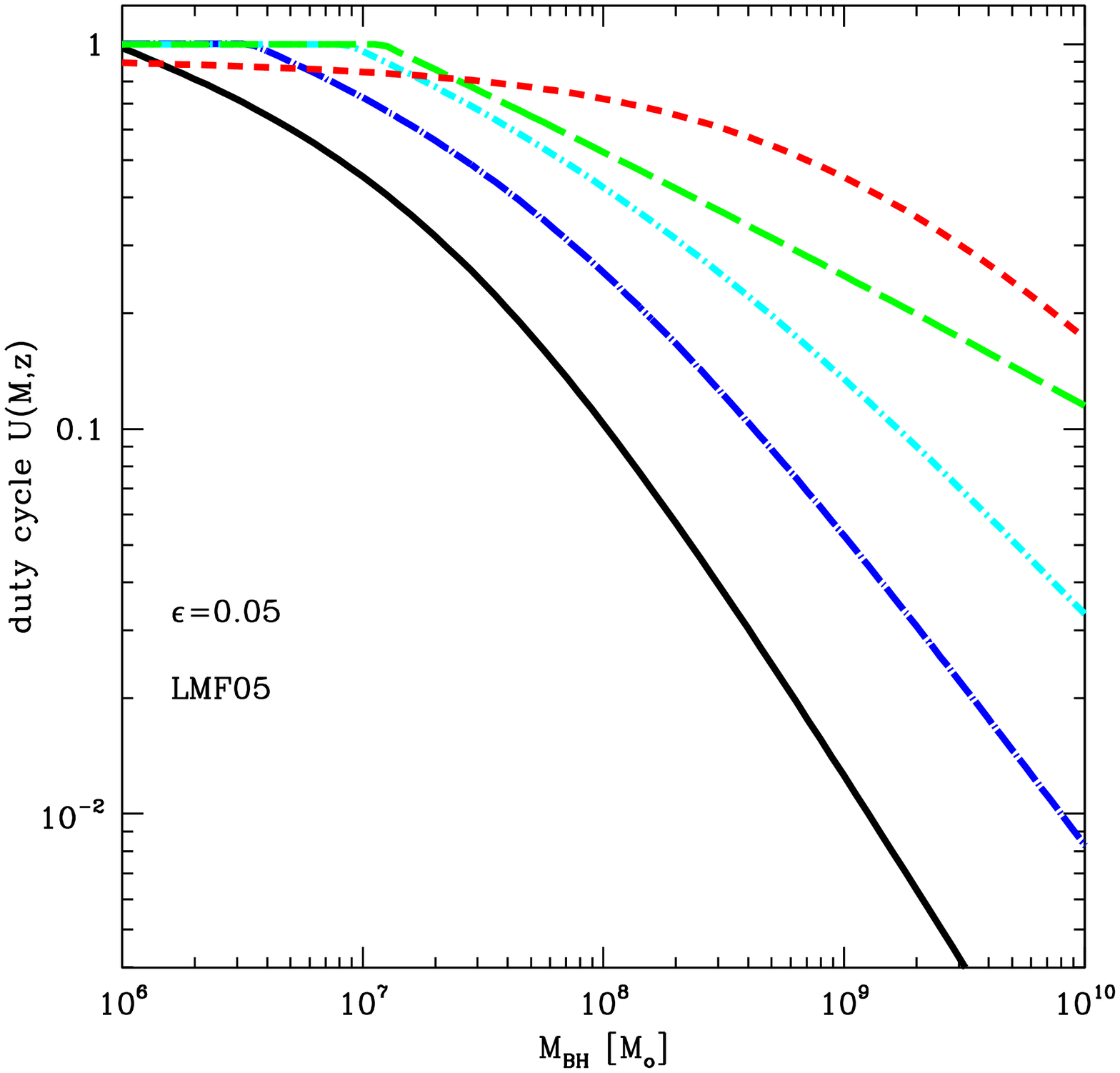}
  \includegraphics[width=6cm]{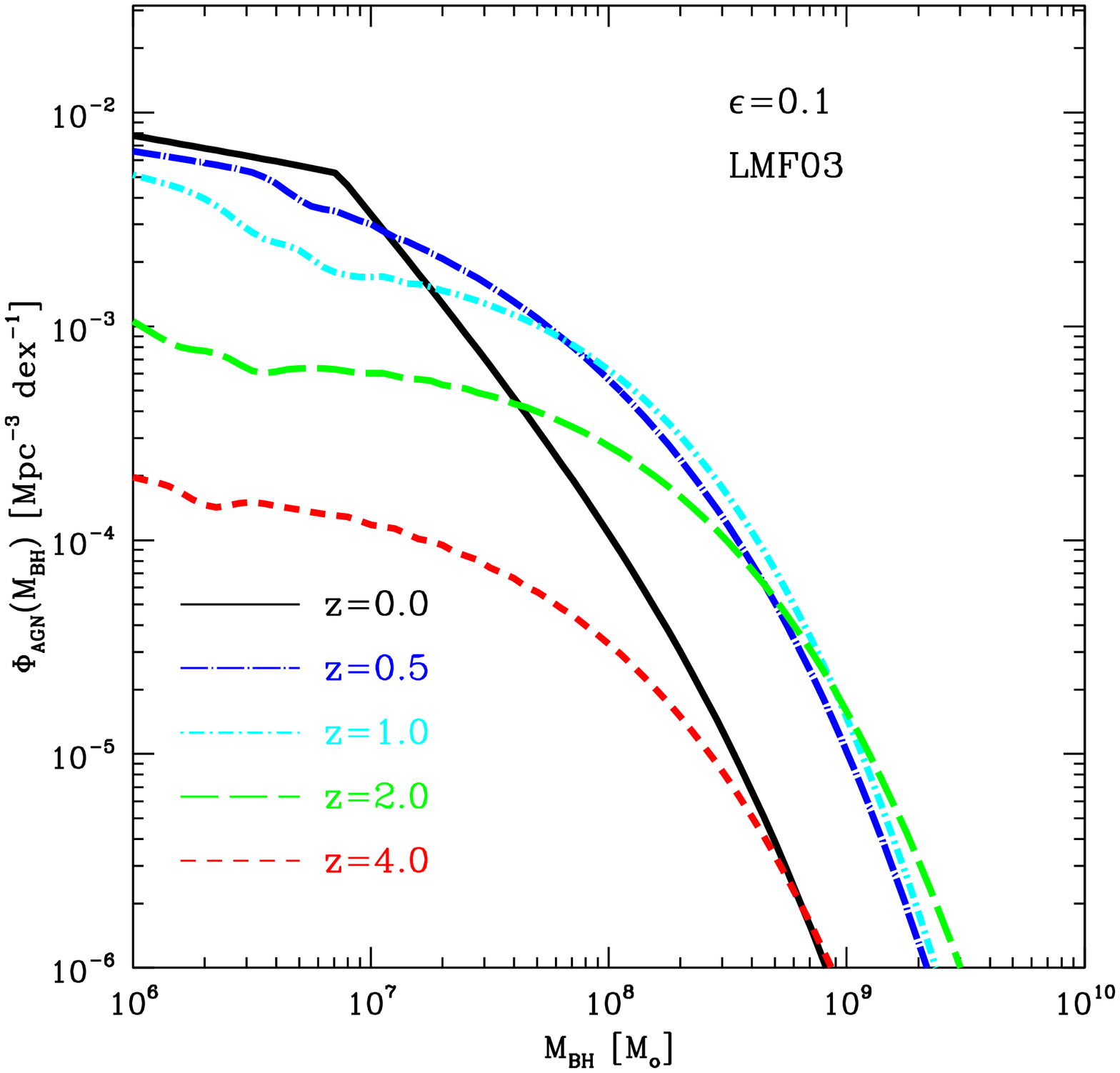}
  \includegraphics[width=6cm]{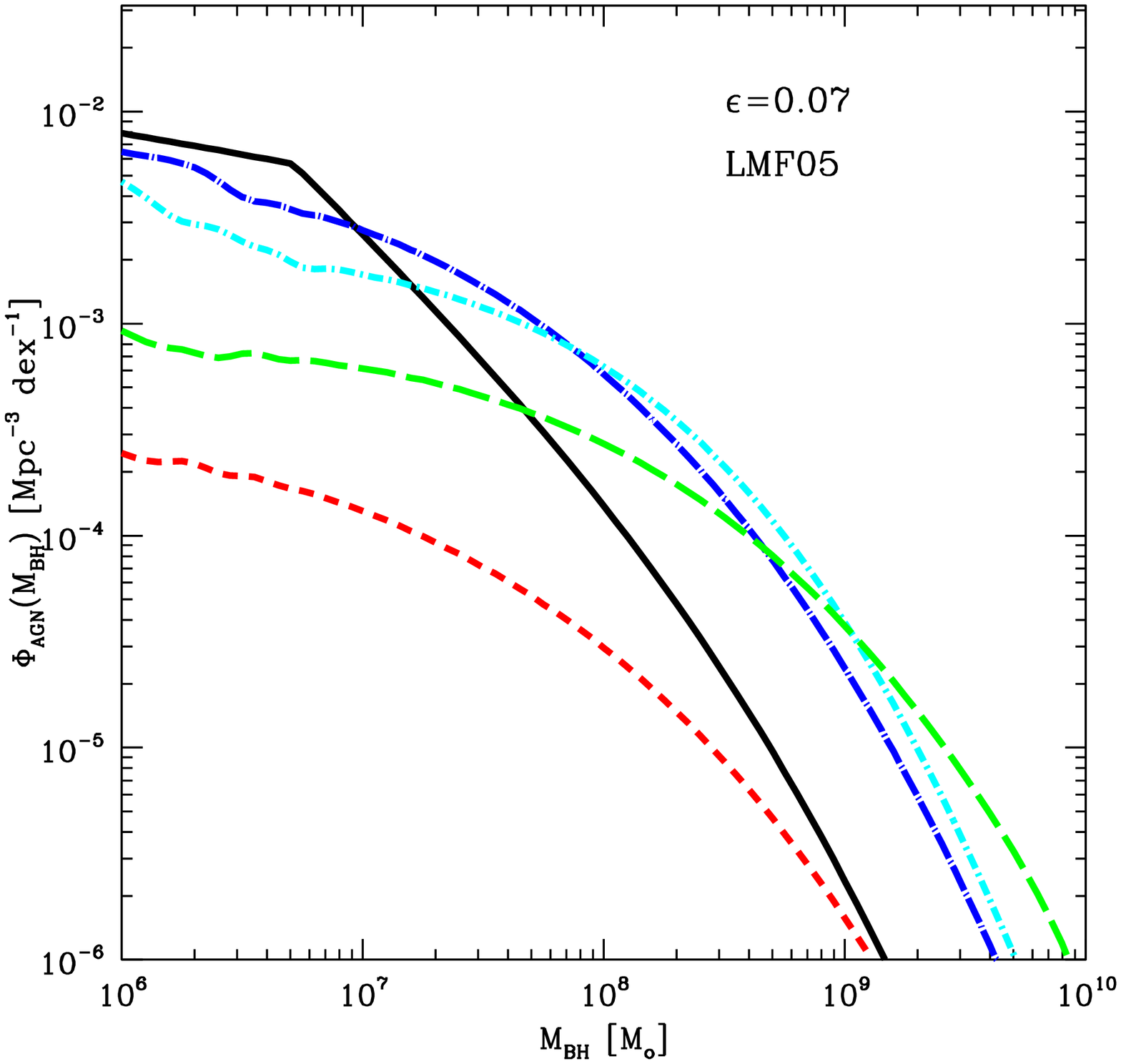}
  \includegraphics[width=6cm]{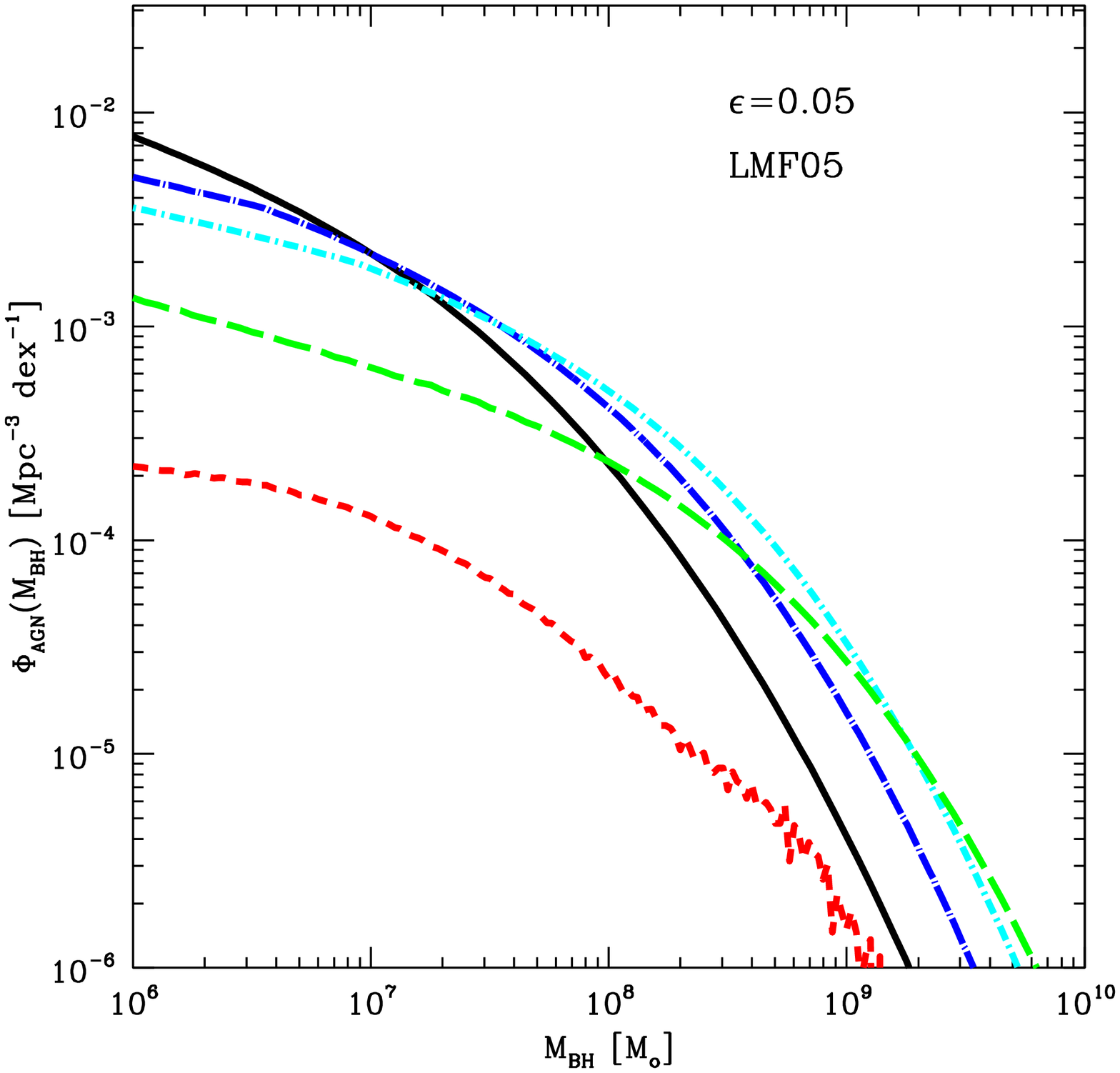}
  \caption{The top panels show the SMBH mass function between redshift
    0 and 4 as predicted by the models with different choices of the
    radiative efficiency and the local mass function. At $z<1,$ the
    BHMF evolves only at $M<10^8\msol$, while number density of very
    massive SMBHs undergoes a strong evolution a higher redshifts. At
    $z>1,$ details of the SMBH cosmic history strongly depend on the
    radiative efficiency. The central panels show the duty cycle
    $U(M,z)$ as predicted by the models. At $z<1,$ the duty cycle is
    $\sim1$ for SMBHs of $M<10^7\msol$ and then decreases with the
    mass. The behavior of the duty cycle is very similar for the
    models with $\epsilon=0.1$ and 0.07, increasing from redshift 0 to
    1, and decreasing at $z>2$. For $\epsilon=0.05,$ it always
    increases and most SMBHs are active at $z\sim4$. The 
      bottom panels show the AGN MF as predicted by the models. This is
    almost insensitive to the choice of the local mass function and to
    the value of the radiative efficiency. The number density of
    active SMBHs with $M\ge10^8\msol$ peaks at redshifts 1--2 and then
    rapidly decreases.}
  \label{s4f2}
\end{figure*}

In Fig.\,\ref{s4f2}, we show the model predictions for the evolution of
the BHMF between redshift 0 and 4. Our results agree with an
anti--hierarchical growth of SMBHs (``cosmic downsizing''), where most
of the low-mass SMBHs were formed later in time than high-mass SMBHs
\citep[e.g.,][]{gra01,ued03,mar04,mer08,sha09}. More quantitatively,
at low redshifts, the mass function evolves only at low/intermediate
masses, and it decreases by approximately 30--60\% from $z=0$ to 1 for objects
of $10^6$--$10^7\msol$. On the contrary, the number density of very
massive SMBHs ($M\ga10^9\msol$) remains practically unchanged up to
$z=1$, and then undergoes a significant evolution.

The details of the SMBH cosmic history depend strongly on the
radiative efficiency. If we compare the results for $\epsilon=0.05$
and 0.07, we note a much faster evolution after redshift 2 if
$\epsilon=0.05$. In this case, the BHMF drops by one order of
magnitude between $z=2$ and 4 in the whole range of masses. Instead,
both the models with $\epsilon=0.1$ and 0.07 predict a similar
moderate evolution of the BHMF: low-mass objects evolve mainly at
redshifts 0-2, while SHBHs with $M>10^8\msol$ evolve between redshifts 1 and
3. In both cases, the MF does not significantly change between $z=3$
and 4, and at these redshifts, it is only a factor $\la10$ lower than
the local BHMF. The main difference between the two models is the
larger number density of very massive SMBHs predicted by the model
with $\epsilon=0.07$; an effect of the different local BHMF employed.

The duty cycle is also shown in Fig.\,\ref{s4f2} and the best-fit
parameters of Eq.\,\ref{s3e6}-\ref{s3e7} are provided in
Table\,\ref{t1}. We can point out a general trend in all three
cases: at $z\la1$ the duty cycle is $\sim1$ at $M\la10^7\msol$ and
then decreases with the SMBH mass; at $z>1$, the duty cycle is less
dependent on $M$  , and for $\epsilon=0.07$ and 0.1 (we note the similar
behavior of $U(M,z)$), it is almost constant at $z\ga2$ and decreasing
with time. On the contrary, for $\epsilon=0.05,$ the duty cycle always
increases with redshift and most SMBHs are active at
$z=4$. In this case, because of the strong evolution of the BHMF, a
high fraction of active SMBHs is required in order to fit the QLF at
high redshifts.

\begin{figure}
  \centering
  \includegraphics[width=\hsize]{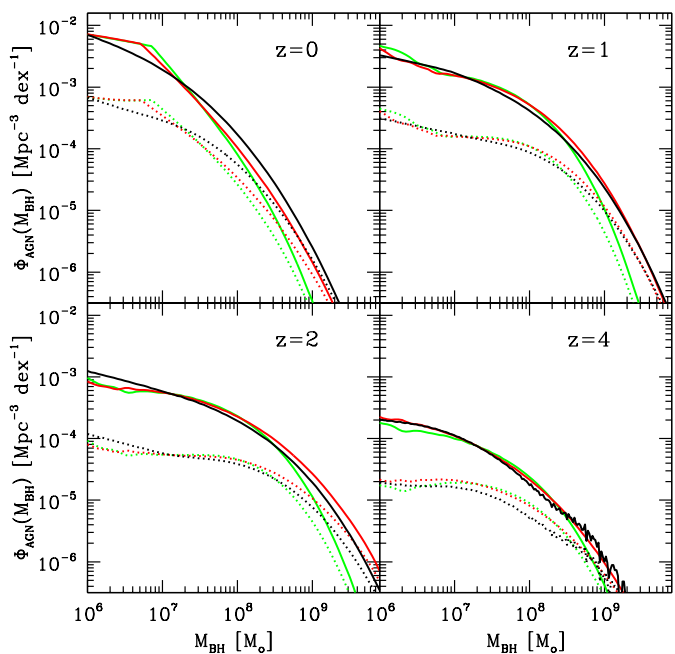}
  \caption{Mass function for type--1 (dotted lines) and type--2 (solid
    lines) AGN according to the models with $\epsilon=0.1$ (green
    lines), $\epsilon=0.07$ (red lines), 0.05 (black lines). The AGN
    mass functions are practically independent of the model in the all
    redshift range. The fraction of type--1 AGN increases with mass.}
  \label{s4f5}
\end{figure}

Given the BHMF and the duty cycle, computing the mass function for
active SMBHs (i.e., the AGN MF) is straightforward. All the models
predict very similar AGN MFs (see Fig.\,\ref{s4f2} and
Fig.\,\ref{s4f5}, where the type--1 and type--2 AGN MFs are given
separately). The only exception is at $M\ga10^9\msol$, where the
  model with $\epsilon=0.1$ predicts a significantly lower number
  density of active SMBHs. However, because it also tends to
  underestimate the observational QLF at high luminosities, we expect
  that the actual MF is larger than predicted by the model in this
  range of masses. This issue may be related to the use of the LMF03
  local BHMF.

From Fig.\,\ref{s4f2} we can note that the number density of
low mass AGN decreases by more than an order of magnitude between
$z=0$ and 4. On the contrary, for massive AGN ($M>10^8\msol$), the
number density peaks at redshifts 1--2 and then rapidly decreases (by
a factor $\sim10$) up to $z=4$. This is a signature of the cosmic
downsizing of AGN. From Fig.\,\ref{s4f5}, we also note the increase of
the fraction of type--1 AGN with the SMBH mass.

It is important to stress that, contrary to what occurs for the BHMF,
the AGN MF is almost insensitive to the choice of the local mass
function and of the value of $\epsilon$. This result can be understood
considering that our method, and standard continuity
equation approaches,  in general, use the QLF as observational constraints, that
depends, through Eq.\,\ref{s2e7}, on the AGN MF, but not on the
BHMF. If the Eddington ratio distribution for SMBHs is known, the QLF
puts strong constraints on the AGN mass function, at least for masses
$M\la10^9\msol$ (more massive SMBHs give a small
contribution to the QLF, as shown in Fig.\,\ref{s4f1}). On the other
hand, the continuity equation method is not able to constrain
details of the SMBH growth history that depend on the model
parameters \citep[see also][]{cap14,vea14}. 
Only a direct measurement of the BHMF at redshifts $z\ga2$ could break
these degeneracies, giving important constraints on the average
radiative efficiency, for example.

\begin{table}
  \caption{Best-fit parameters of the duty cycle (see
    Eq.\,\ref{s3e6}--\ref{s3e7}) for the different values of
    the radiative efficiency.}
\label{t1}
\centering
\begin{tabular}{cccccc}
\hline
$\epsilon$ & $X$ & $a_X$ & $b_X$ & $c_X$ & $d_X$ \\
\hline
0.1 & A             & 0.956 & -1.337 & 0.383 & -0.0450 \\
     & $\alpha_l$ & 0.250 & -0.211 & 0.0434 & -1.82e-04 \\
     & $\alpha_k$ & 1.159 & -1.571 & 1.354 & -0.0256 \\
     & $M_0$       & 4.884 & 4.648 & -1.284 & 0.109 \\
\hline
0.07 & A             & 0.634 & -0.803 & 0.151 & -0.0122 \\
      & $\alpha_l$ & 0.944 & -0.706 & 0.194 & 0.438 \\
      & $\alpha_k$ & 0.0456 & 0.149 & -0.132 & 0.0259 \\
      & $M_0$       & 4.965 & 4.913 & -1.545 & 0.183 \\
\hline
0.05 & A             & -0.0104 & 0.309 & -0.0659 & -3.70e-03 \\
       & $\alpha_l$ & 1.012 & -0.335 & 0.0123 & 2.06e-03 \\
       & $\alpha_k$ & 0.195 & -0.0570 & 0.0348 & 2.10e-03 \\
       & $M_0$       & 7.389 & 0.714 & -0.0373 & -8.98e-03 \\
\hline
\end{tabular}
\end{table}

\begin{figure}
  \centering
  \includegraphics[width=\hsize]{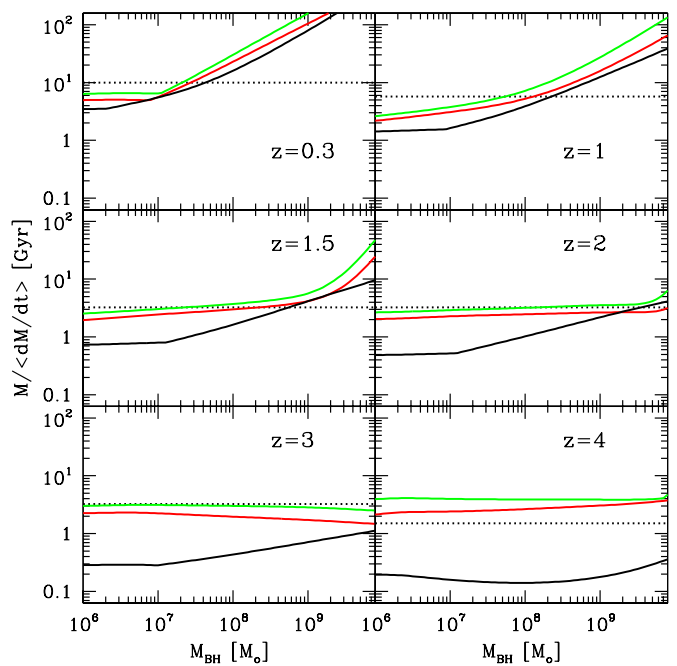}
  \caption{Redshift evolution of the growth time as a function of SMBH
    mass. In each panel, the horizontal dotted line marks the age of
    the Universe at that redshift. Lines indicate the different values
    of $\epsilon$: black, red and green lines are for $\epsilon=0.05$,
    0.07 and 0.1, respectively. The main epoch of growth is at
    $1<z<3$. At later time, only low--mass SMBHs can be actually
    growing in mass, while at earlier time only in the model with
    $\epsilon=0.05$ the growth time is much lower than the age of the
    Universe.}
  \label{s4f6}
\end{figure}

\begin{figure}
  \centering
  \includegraphics[width=\hsize]{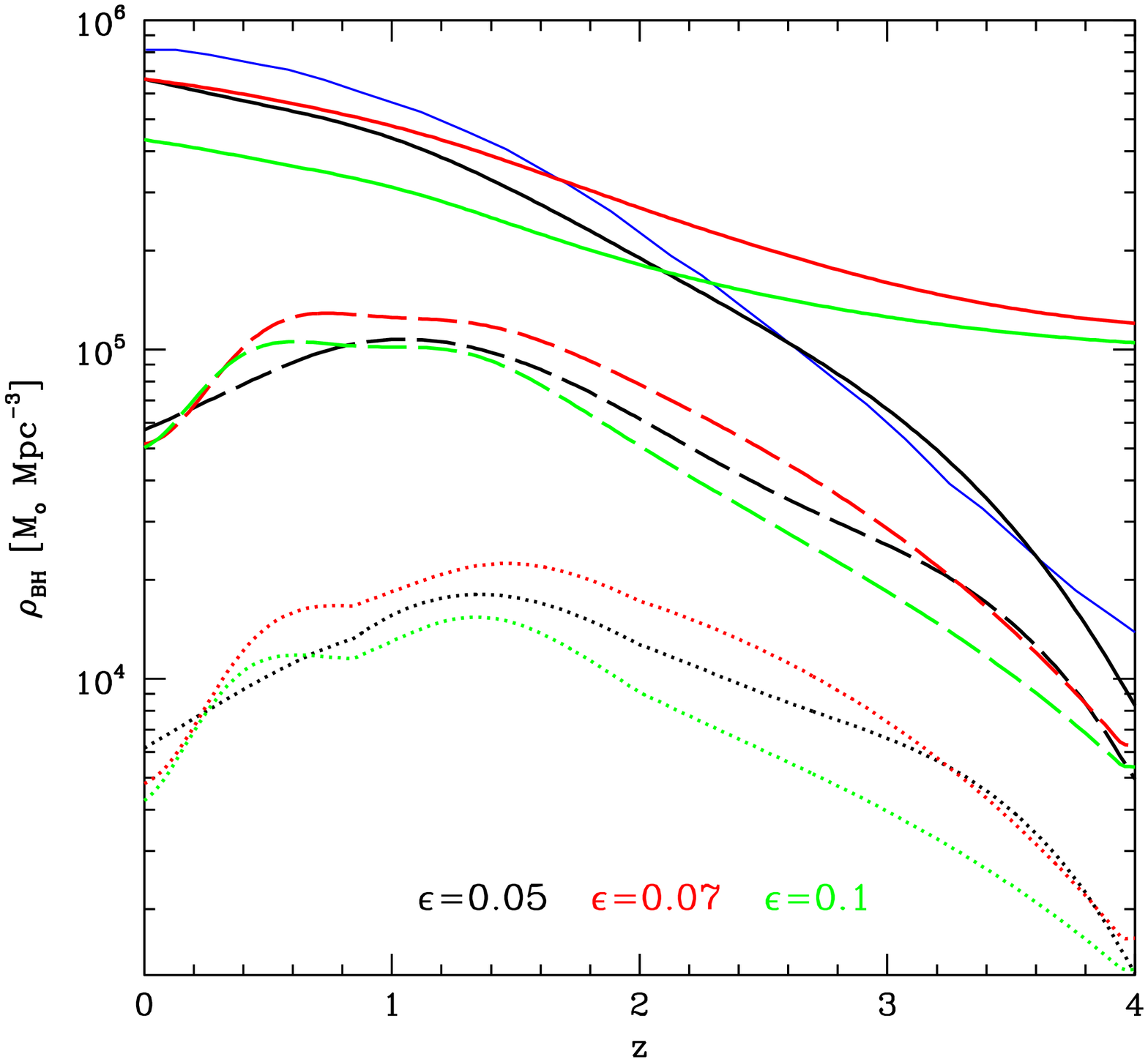}
  \caption{Mass density of SMBHs (solid lines), active SMBHs
    (dashed lines), and type--1 AGN (dotted lines) for the three values
    of the radiative efficiency indicated in the plot. The thin blue
    solid line corresponds to results from \citet{ued14}. If
    $\epsilon=0.1$ or 0.07, the evolution of the mass density is relatively
    modest. On the contrary, if $\epsilon=0.05$, the mass density at
    $z=4$ is two orders of magnitude lower than the local one. The AGN
    mass density has similar evolution for the three cases, with a
    peak at approximately $z=0.5$--1.5.}
  \label{s4f7}
\end{figure}

A quantity that is useful for gaining information on the typical
accretion rate of SMBHs is the growth time, that is, the ratio
$M/\arate$ \citep{mer08}. Fig.\,\ref{s4f6} shows the redshift
evolution of the growth time as a function of SMBH mass. In each
panel, we draw the age of the Universe at that time as a
reference. SMBHs with growth time longer than the age of the Universe
are not experiencing a major growth phase, which must have necessarily
happened in the past. On the contrary, objects with growth times
shorter than that are actively growing. Looking at Fig.\,\ref{s4f6},
we see that only small mass SMBHs ($M\la10^7\msol$) can be actually
growing in mass at low redshifts (see also Fig.\,\ref{s3f4}). This is
expected according to the downsizing behavior of SMBHs, in which only
low-mass SMBHs are still accreting mass and growing in the local
universe or at low redshift; in agreement with observational
constraints \citep{hec04,gre07,gou10,sch10}. The main epoch of growth
for SMBHs is around redshift 1.5-3. At these redshifts, the growth
time is almost independent of $M$ and below the age of the
Universe. At $z>3,$ the growth time becomes typically larger than the
age of the Universe, except for $\epsilon=0.05$. In this case, the
growth time steadily decreases and SMBHs are active and growing
independently of their mass, up to at least redshift 4.

Fig.\,\ref{s4f7} illustrates the evolution of the integrated SMBH mass
density, $\rho_{BH}=\int \log(M)M\Phi_{BH}(M)$, as predicted by the
models. The interesting feature is the very different time evolution
of $\rho_{BH}$ between the model with $\epsilon=0.05$ and 0.07/0.1. In
the latter, the evolution is modest, decreasing by a factor approximately 5
from redshift 0 to 4. On the contrary, for $\epsilon=0.05,$ the mass
density is lower than $10^4\msol$\,Mpc$^{-3}$ at $z=4$, that is, approximately
two orders of magnitude below the local density. This behavior is in
very good agreement with observational findings from \citet{ued14}
  obtained from a compilation of AGN X-ray luminosity surveys
assuming a radiative efficiency of 0.05.

The different evolution of the BHMF, as predicted by our models, has
important implications on models for the seed population of SMBHs
\citep[see, e.g.,][]{vol08,vol10}. Our results with large radiative
efficiencies ($\epsilon\ge0.07$) indicate that a large fraction of
SMBHs were already formed at $z>4$. This implies very massive primordial
black hole seeds, as expected from the direct collapse of supermassive
stars \citep[e.g.,][]{kou04,beg06,lod07,dev09}. The case with
$\epsilon=0.05$ is instead more compatible with models in which
primordial black holes originate from stellar mass progenitors
\citep[remnants of the first, Population III, stars; see
e.g.,][]{abe00,bro02} and that predict negligible SMBH mass density at
high redshifs.

If we consider active SMBHs, Fig.\,\ref{s4f7} shows a peak in the mass
density of type--2 and type--1 AGN at redshifts of approximately 0.5-1.5. The
peak is more pronounced in type--1 AGN due to the increase of their
fraction with luminosity and redshift. As expected, the evolution of
the AGN mass density is almost independent of the model. The mass
density is somewhat lower for $\epsilon=0.1$, an effect of the sharper
drop of the mass function at $M\ga10^9\msol$, observed in
Fig.\,\ref{s4f5}.

\subsection{Comparison with observations}
\label{s4s4}

Observational estimates of the quasar mass function and of the
ERDF are a promising test for
model predictions. Nevertheless, direct comparison between model
and observations is not trivial due to selection effects in quasar
samples. A standard approach to compute the MF and ERDF is based on
the $1/V_{max}$ estimator with the same volume weights as for the QLF
\citep[e.g.,][]{wan06,ves08,ves09,sch10,nob12}. This method has shown
to suffer from severe incompleteness due to active SMBHs below the
flux limit of the survey that are not taken into account \citep[see
discussion in][]{kel09,sch10,sch15}. Uncertainties and scatter in the
relation to estimate SMBH masses have also to be considered for a
proper MF determination. Several studies have estimated the MF and
ERDF for type--1 AGN, employing statistical methods to properly
account for the survey selection function and the uncertainties in SMBH
mass estimates \citep{kel09,sch10,she12,nob12,kel13,sch15}. Their
results are model dependent because they use analytic functions to
describe the bivariate distribution function of $M$ and $\lambda$.

\begin{figure*}
  \centering
  \includegraphics[width=9cm]{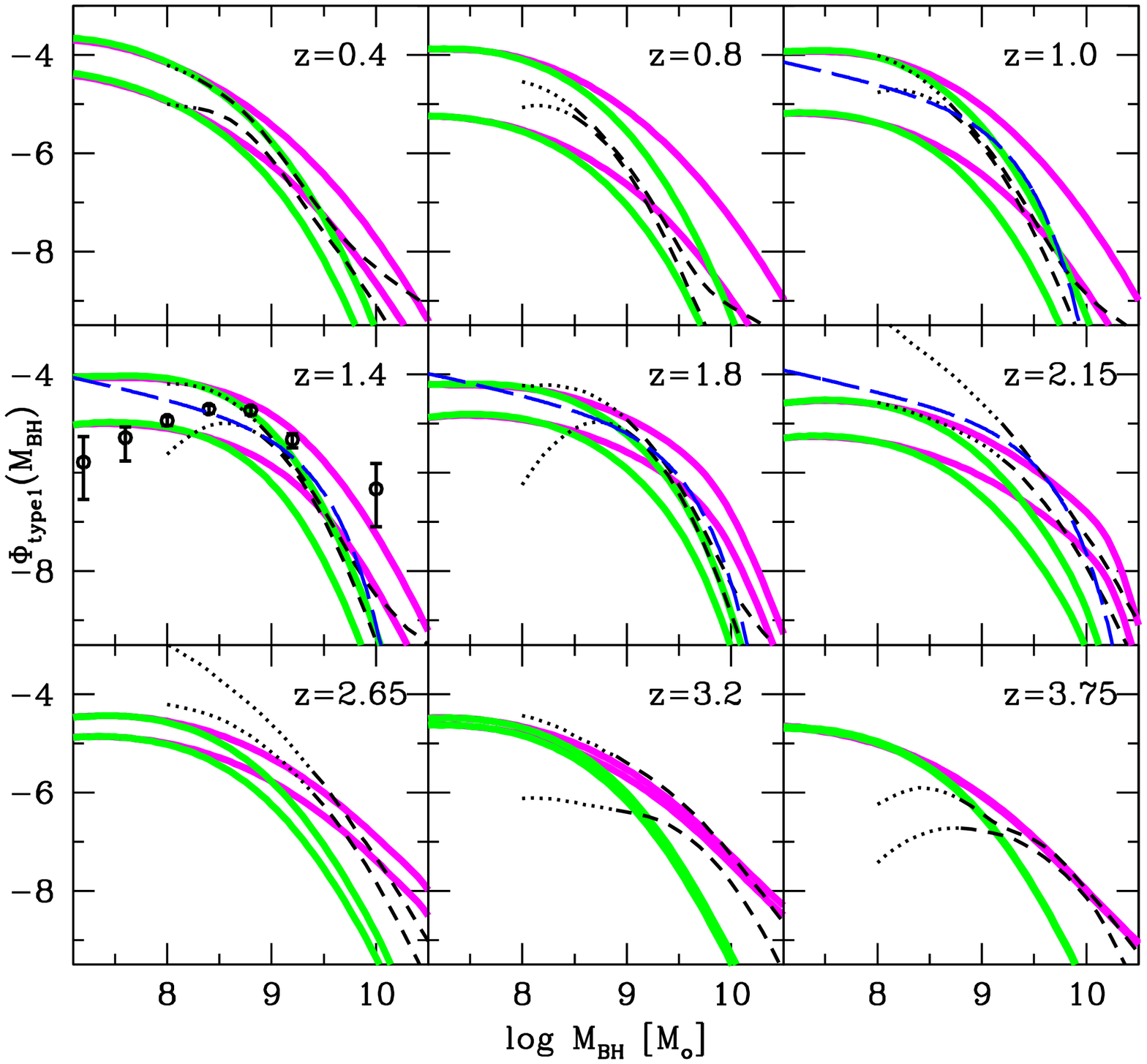}
  \includegraphics[width=9cm]{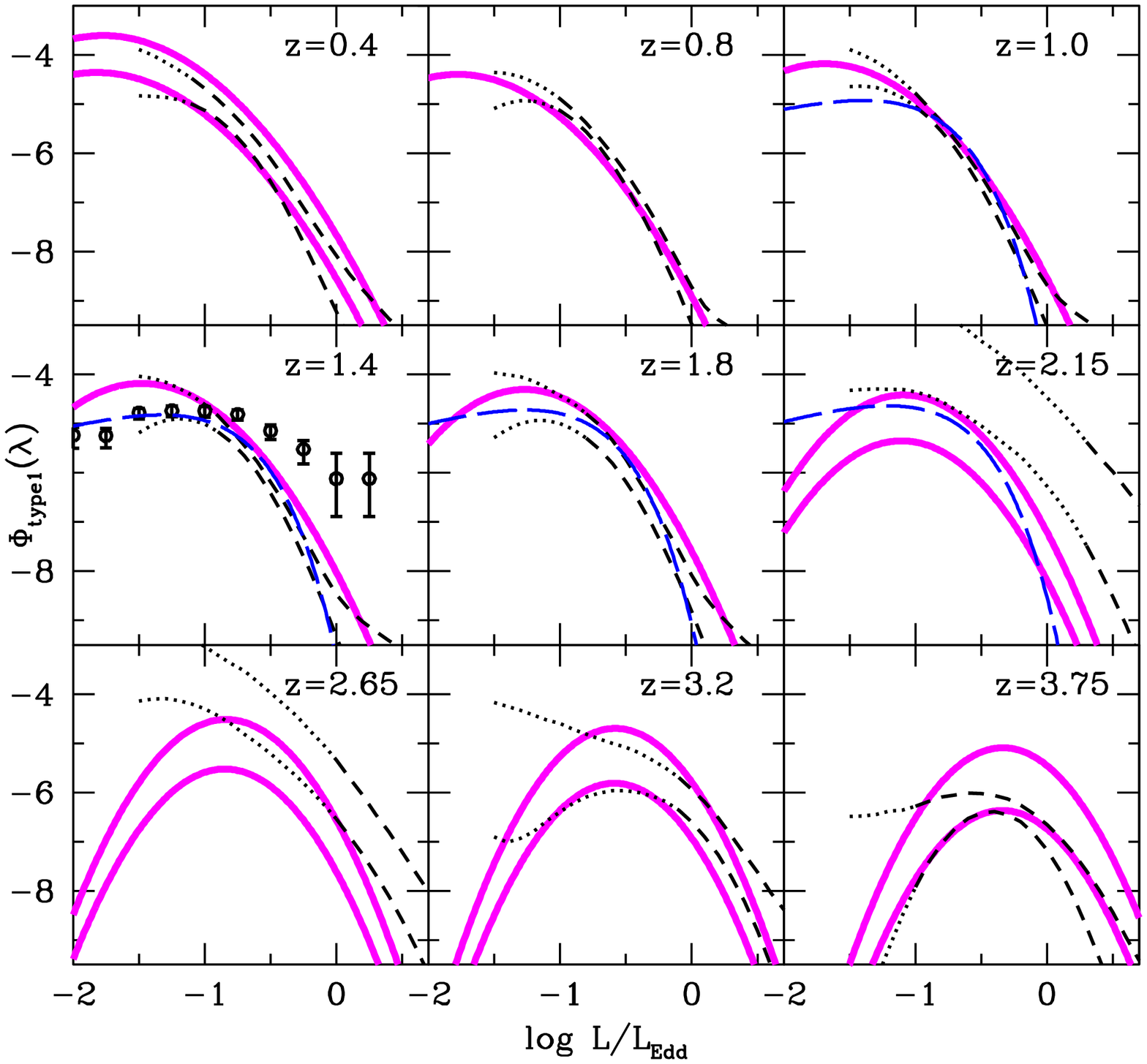}
  \caption{The left panel shows the MFs for type--1 AGN computed from
    the model and compared with observational estimates. Solid thick
    lines are for the model with $\epsilon=0.07$ (magenta lines) and
    $\epsilon=0.1$ (green lines), assuming the values of
    $\lambda_{min}$ reported in Table\,\ref{t2}. Observational
    constraints are from \citet{sch15} (at $1\le z\le2.15$; blue long
    dashed lines), \citet{nob12} (at $z=1.4$; black open points) and
    \citet{kel13} (black dashed/dotted lines; the two lines define the
    region of 68\% probability, and the dotted lines denote that the
    completeness for the SDSS sample is below 10\%). The right
      panel shows the ERDF for type--1 AGN, as in the left panel. Here we
    plot only the model with $\epsilon=0.07$.}
  \label{s4f10}
\end{figure*}

\citet{kel13} determined the MF and the ERDF for a sample of type--1
AGN from the SDSS at redshifts $0.4<z<5$. They found that the sample
becomes significantly incomplete ($\la10$\%) at $M\la3\times10^8\msol$
or $\lambda\la0.07$, with some variation with redshift. In
Fig.\,\ref{s4f10}, we plot their results and compare them with
predictions of our models. At a fixed redshift, we compute the MF of
type--1 AGN as:
\beq
\Phi_{AGN}^{type1}(M)=\Phi_{AGN}(M)\,a_n(M)
\int_{\log{\lambda_{min}}}^1 d\log\lambda\,
f_{uno}(L)P_1(\lambda)
\label{s4e1}
,\eeq
and the ERDF as:
\beq
\Phi_{AGN}^{type1}(\lambda)=P_1(\lambda)\int_{\log M_{min}}^{11}
d\log M\,a_n(M)f_{uno}(L)\,\Phi_{AGN}(M)\,,
\label{s4e2}
\eeq
where $f_{uno}(L)$ is the fraction of type--1 AGN with luminosity $L$,
and $M_{min}$ ($\lambda_{min}$) is the minimum SMBH mass (Eddington
ratio) associated to the flux--limited survey used in
\citet{kel13}. The actual values of $M_{min}$ and $\lambda_{min}$ are
not well determined and we choose two values that should represent
upper and lower limits for them (see Table\,\ref{t2}).

Fig.\,\ref{s4f10} shows our predictions for the models with
$\epsilon=0.1$ and 0.07 (the results are similar for $\epsilon=0.05$
and 0.07 models). At redshifts $z<2,$ the MF is relatively highly dependent on the
choice of $\lambda_{min}$ due to the broad shape of the ERDF that
peaks at $0.01\le\lambda<0.1$. The comparison with observations is
more interesting at high redshifts where the choice of $\lambda_{min}$
is less important. In general, the models are reasonably consistent
with observational estimates of \citet{kel13}. The case with
$\epsilon=0.1$ better reproduces the shape of the observational MF at
low redshifts, but fails at high redshifts where it strongly
underestimates the observations for $M\ga10^9\msol$. Viceversa, models
with $\epsilon=0.05$/0.07 fit the data at $z\ga2$ relatively well, but at
low redshifts, the predicted MF seems to decrease too slowly with the
mass and gives an excess of high-mass type--1 AGN. We must note,
however, that SMBHs with $M\ga5\times10^9\msol$ give a negligible
contribution to the QLF at $z<2$ (see Fig.\,\ref{s4f1}), and the AGN
MF is therefore poorly constrained by the analysis at these
masses. This argument does not work for the discrepancy observed in
the $\epsilon=0.1$ model: the fast decline of the MF at high redshifts
cannot be reconciled with the observations even assuming duty cycles
equal 1.

Concerning the ERDF, we reiterate that the model uses a log-normal
function that fits the shape of the Eddington ratio distribution
determined by \citet{kel13} at the different redshifts. Therefore, in
Fig.\,\ref{s4f10}, we simply verify whether or not the amplitude of the ERDF
computed by Eq.\,\ref{s4e2} is consistent with observational
results. The agreement is good with $M_{min}=10^8\msol$, apart from
the first and last redshift bin in which $M_{min}=10^7$ and
$10^9\msol$, respectively, provide a better fit to the data. The only
discrepancies we observe are at $z=2.15$ and 2.65, where our
predictions are significantly lower than the data. However, \citet{kel13}
questioned the reliability of their estimates at these redshift bins
because they found an apparent discontinuity across $z\sim2$ in the
number densities of type--1 AGN radiating at $\lambda\ga0.05$. They
attributed the discontinuity to systematic errors in the
incompleteness correction. This, however, could  also be due to the different
mass estimator used before and after $z\sim2$ \citep{sch15}.

\begin{table}
\caption{Values of $M_{min}$ and $\lambda_{min}$ used in
  Eqs.\,\ref{s4e1} and \ref{s4e2} to compute AGN MF and ERDF for the
  comparison with \citet{kel13} estimates.}
\label{t2}
\centering
\begin{tabular}{ccc}
\hline
Redshift & $\log M_{min} [\msol]$ & $\lambda_{min}$ \\
\hline
0.4 & 7.,\,8. & 0.01,\,0.05 \\
(0.4,2) & 8. & 0.01,\,0.1 \\
>2 & 8.,\,9. & 0.07,\,0.2 \\
\hline
\end{tabular}
\end{table}

\begin{figure}
  \centering
 \includegraphics[width=\hsize]{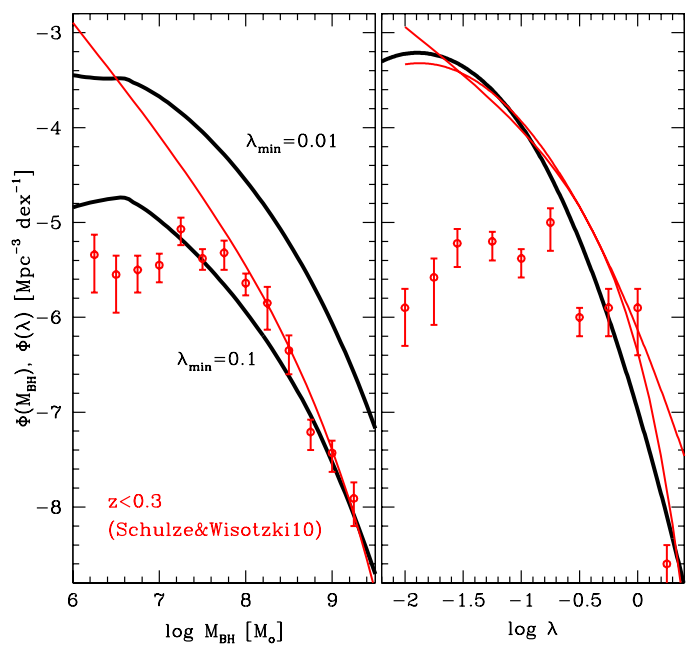}
  \caption{The local ($z=0.1$) BHMF (left panel) and ERDF (right
    panel) for type--1 AGN predicted by the model with $\epsilon=0.07$
    (solid black thick lines). The BHMF is computed taking
    $\lambda_{min}=0.01$ and 0.1. The ERDF is computed for
    $M_{min}=10^6\msol$. Red points are the distribution functions
    derived by \citet{sch10} directly from data, while thin red lines
    are their best fits, taking the sample selection function into
    account.}
  \label{s4f11}
\end{figure}

Our predictions on the ERDF at $z\sim2$ are instead consistent with
results from \citet[][blue lines in Fig.\,\ref{s4f10}]{sch15}. They
combined large area SDSS data with two deep, small-area surveys (VVDS
and zCOSMOS) to cover a wide range of luminosities at redshifts $1\la
z\la2$. They use a maximum likelihood approach to fit a parametric
bivariate distribution function in the intervals of $-2<\log\lambda<1$
and $7<\log(M/\msol<11$. Their estimates agree well with the
\citet{kel13} MF and ERDF at redshift $z<2$. In Fig.\,\ref{s4f10}, we
also consider the results from \citet{nob12}, that used the Subaru
XMM--{\it Newton} Deep Survey (SXDS) to determine the MF and ERDF at
$z\sim1.4$. Their sample is X-ray selected and extends significantly
deeper than SDSS, over an area of $\sim1.0$\,deg$^2$. Their MF is in
agreement with the other estimates in the mass range 8--9.5$\msol$,
while they show a turnover at lower masses not confirmed by the other
data. Some discrepancies are also observed in the ERDF that gives a
much larger probability for high accretion SMBHs.

Finally, in Fig.\,\ref{s4f11}, we consider the local MF and ERDF for
type--1 AGN computed by \citet{sch10}. They used a sample of local (z
< 0.3) broad line AGN from the Hamburg/ESO Survey. In the Figure, we
report the MF and ERDF directly determined from the data (red points)
and after the incompleteness correction through a maximum likelihood
approach (red lines). The model seems to be in better agreement with
the observational MF before the incompleteness correction (it requires
$\lambda_{min}\simeq0.1$). On the contrary, the predicted ERDF
is only consistent with data after the incompleteness correction.

\section{Conclusions}
\label{s5}

In this paper, we study the time evolution of the mass function
of SMBHs and of their active population (i.e., SMBHs with
$\lambda\ge10^{-4}$) by the continuity equation, backwards in time
from the present to redshift 4. In our approach, we distinguish active SMBHs
between type-1 and type-2 AGN. The Eddington ratio distribution of
the two classes is chosen on the basis of recent observational
estimates, assuming a log-normal distribution and a truncated power
law for type-1 and type-2 AGN, respectively. The duty cycle of
SMBHs, as a function of redshift and mass, is instead determined from
the best fit of the observational quasar luminosity functions of
\citet{hop07}.


We also investigate the dependence of the SMBH and AGN evolution
on the main assumptions/inputs employed in the analysis. These are:
(1) the value of the average radiative efficiency of SMBHs, which is
the only free parameter of the model; (2) the local SMBH mass
function; and (3) the Eddington ratio distribution for type--1 and type--2
AGN. Below we summarize and discuss our results.

\begin{itemize}

\item The evolution of the BHMF, especially at $z\ga2$, is very
  sensitive to the value of the average radiative efficiency. This is
  clearly shown by comparing the results using $\epsilon=0.07$
  and 0.05: with the lower radiative efficiency, the evolution of the
  BHMF is significantly stronger at $z\ga2,$ and at $z=4$ the mass
  function is one order of magnitude lower than that with
  $\epsilon=0.07$. For larger radiative efficiencies, we find very
  little evolution in the MF, especially at low masses, and the number
  density of SMBHs (of $M\la10^9\msol$) increased by only a factor
of approximately $4$ (2) from $z=4$ to 0 if $\epsilon=0.1$ (0.15). 

\item Radiative efficiencies much larger than 0.1 seem to be discarded
  by our model. In this case, in fact, the BHMF is almost constant
  over time, implying that most SMBHs we observe at low redshifts were
  already formed before redshift 4. For large radiative efficiences, a
  significant evolution in time of the BHMF could only be expected if
  the local BHMF was substantially overestimated, and in this case
  larger duty cycles at low redshifts would be needed.

\item Independently of the parameters of the model, we confirm an
  anti--hierarchical growth of SMBHs. Black holes of $M\ga10^9\msol$
  stop forming at redshift 1 in all our models, while lower-mass
  SMBHs typically grow later or keep growing in all the
  redshift range according to the model. Anti-hierarchical
  behavior can also be observed in the duty cycle; at low redshift, this
  is close to 1 for objects of $M\la10^7\msol$ and 
  decreases rapidly with the mass.

\item Results for $\epsilon\ge0.07$ and $\epsilon\sim0.05$ imply quite
  different scenarios for the SMBH evolution at high redshifts. In the
  first case, a large number density of SMBHs, most of them quiescent,
  are already in place at $z\ga4$ (the mass density is approximately
  $10^5\msol$\,Mpc$^{-3}$). On the other hand, the model with
  $\epsilon=0.05$ predicts a small number density of SMBHs at high
  redshifts ($\rho_{BH}<10^4\msol$\,Mpc$^{-3}$ $z\sim4$). The duty
  cycle is steadily increasing with redshift and most SMBHs are
  active at $z\simeq4$. 

\item The evolution of active SMBHs also shows an anti--hierarchical
  behavior. The MF of low-mass AGN steadily increases with time,
  while for intermediate/high-mass AGN, we find a peak in the number
  density between redshift 1 and 2. As an example, the number density of
  $M\sim10^9\msol$ AGN peaks around $z=1.5$ and it is approximately eight times
  larger than it is at present (and than at $z=3.4$, whose number density is equal
  to the local one).

\item Information on the actual Eddington ratio distribution of SMBHs
  is still incomplete and uncertain. We have tested our results
  against different choices of the Eddington ratio distribution for
  type--1 and type--2 AGN. In general, we observe a modest impact on
  the evolution of the AGN MF, compatible with the uncertainty of the
  model.

\end{itemize}

\begin{figure}
  \centering
 \includegraphics[width=\hsize]{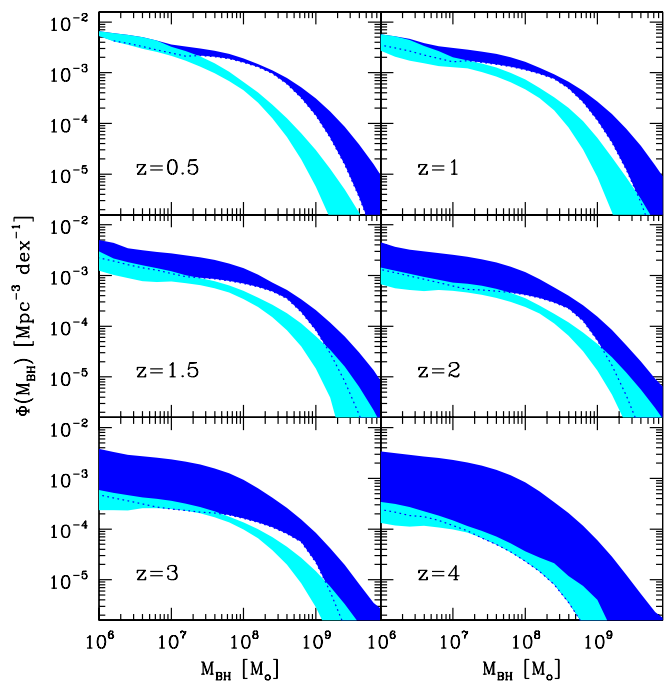}
 \caption{The uncertainty on the predicted MF for SMBHs (blue shaded
   areas) and for AGN (cyan shaded areas) at different redshifts. This
   is obtained by combining the results and the related uncertainties
   from the models with $\epsilon=0.05,$ 0.07, and 0.1. Dotted lines
   are the lower limits for the BHMF. We note the strong
   constraints imposed by the models on the evolution of AGN, compared
   to the large uncertainty in the BHMF at high redshifts.}
  \label{s5f1}
\end{figure}

The main results of the paper can be summarized by
Fig.\,\ref{s5f1}. This provides a realistic estimate of the
uncertainties on the SMBH and AGN mass function that arise from a
continuity equation approach, on the basis of current knowledge
(on, e.g., $\epsilon$, the local BHMF and QLF). These uncertainties
are obtained by combining the results from the three best models
discussed in the paper. We conclude that robust and strict predictions
can be provided on the evolution of the mass function for type-1 and
type-2 AGN. The AGN MF is mainly determined by the QLF, with a small
dependence on the choice of the radiative efficiency or of the local
BHMF. The uncertainty only increases for very massive SMBHs, whose
contribution to the observational QLF is small or negligible. On the
other hand, we are less predictive of the evolution of the BHMF at
high redshifts, which is very sensitive to model parameters.
  
This approach is complementary to other techniques that attempt to
constrain the cosmic evolution of SMBHs and AGN, from analytical
forward models to ab-initio semi-analytical models and cosmological
simulations.  Our framework, once it has been calibrated to match the
global properties of the {observed} population, as done in this
paper, can be used as a starting point to investigate additional
properties and make further predictions without tuning the set
parameters.

The uncertainties in the (astro)physics of SMBHs and the lack of
observational constraints, especially at high redshift and faint AGN
luminosities, still allow relative freedom in choosing free parameters
in analytical models and ``sub-grid" physics.  With the advent of
future space missions, such as {\it Athena} and JSWT, dedicated to
faint, high-redshift sources, we will be able to reduce the freedom in
models, and advance our theoretical understanding.

\begin{acknowledgements}

  MT thanks Martin Kunz for the helpful discussions on MCMC methods
  and their applications, and for providing the MCMC code used in this
  paper. MT also acknowledge support by the Swiss National Science
  Foundation. MV acknowledges funding from the European Research
  Council under the European Community's Seventh Framework Programme
  (FP7/2007-2013 Grant Agreement no.\ 614199, project ``BLACK''). Part
  of the analysis was performed on the BAOBAB cluster at the
  University of Geneva.

\end{acknowledgements}


\begin{thebibliography}{}

\bibitem[Abel et al.(2000)]{abe00} 
Abel, T., Bryan, G.~L., \& Norman, M.~L.\ 2000, \apj, 540, 39 

\bibitem[Aird et al.(2012)]{air12} 
Aird, J., Coil, A.~L., Moustakas, J., et al.\ 2012, \apj, 746, 90 

\bibitem[Aversa et al.(2015)]{ave15} 
Aversa, R., Lapi, A., de Zotti, G., Shankar, F., \& Danese, L.\ 2015, arXiv:1507.07318 

\bibitem[Babi{\'c} et al.(2007)]{bab07} 
Babi{\'c}, A., Miller, L., Jarvis, M.~J., et al.\ 2007, \aap, 474, 755 

\bibitem[Begelman et al.(2006)]{beg06} 
Begelman, M.~C., Volonteri, M., \& Rees, M.~J.\ 2006, \mnras, 370, 289 

\bibitem[Berti \& Volonteri(2008)]{ber08} 
Berti, E., \& Volonteri, M.\ 2008, \apj, 684, 822 

\bibitem[Bongiorno et al.(2007)]{bon07} 
Bongiorno, A., Zamorani, G., Gavignaud, I., et al.\ 2007, \aap, 472, 443 

\bibitem[Bongiorno et al.(2012)]{bon12} 
Bongiorno, A., Merloni, A., Brusa, M., et al.\ 2012, \mnras, 427, 3103 

\bibitem[\protect\citeauthoryear{{Booth} \& {Schaye}}{{Booth} \&
  {Schaye}}{2009}]{Booth2009}
{Booth} C.~M.,  {Schaye} J.,  2009, \mnras, 398, 53

\bibitem[\protect\citeauthoryear{{Bower}, {Benson}, {Malbon}, {Helly}, {Frenk},
  {Baugh}, {Cole} \& {Lacey}}{{Bower} et~al.}{2006}]{2006MNRAS.370..645B}
{Bower} R.~G.,  {Benson} A.~J.,  {Malbon} R.,  {Helly} J.~C.,  {Frenk} C.~S.,
  {Baugh} C.~M.,  {Cole} S.,    {Lacey} C.~G.,  2006, \mnras, 370, 645

\bibitem[Bromm et al.(2002)]{bro02} 
Bromm, V., Coppi, P.~S., \& Larson, R.~B.\ 2002, \apj, 564, 23 

\bibitem[Cao(2010)]{cao10} 
Cao, X.\ 2010, \apj, 725, 388 

\bibitem[Cao \& Li(2008)]{cao08} 
Cao, X., \& Li, F.\ 2008, \mnras, 390, 561 

\bibitem[Caplar et al.(2014)]{cap14} 
Caplar, N., Lilly, S., \& Trakhtenbrot, B.\ 2014, arXiv:1411.3719 

\bibitem[Cavaliere et al.(1971)]{cav71} 
Cavaliere, A., Morrison, P., \& Wood, K.\ 1971, \apj, 170, 223

\bibitem[\protect\citeauthoryear{{Cavaliere} \& {Vittorini}}{{Cavaliere} \&
  {Vittorini}}{2000}]{2000ApJ...543..599C}
{Cavaliere} A.,  {Vittorini} V.,  2000, \apj, 543, 599

\bibitem[\protect\citeauthoryear{{Cattaneo}}{{Cattaneo}}{2001}]{Cattaneo2001}
{Cattaneo} A.,  2001, MNRAS, 324, 128

\bibitem[\protect\citeauthoryear{{Cattaneo}, {Blaizot}, {Devriendt} \&
  {Guiderdoni}}{{Cattaneo} et~al.}{2005}]{Cattaneo2005}
{Cattaneo} A.,  {Blaizot} J.,  {Devriendt} J.,    {Guiderdoni} B.,  2005,
  MNRAS, 364, 407

\bibitem[Croom et al.(2009)]{cro09} 
Croom, S.~M., Richards, G.~T., Shanks, T., et al.\ 2009, \mnras, 399, 1755 

\bibitem[\protect\citeauthoryear{{Croton} et~al.,}{{Croton}
  et~al.}{2006}]{Croton2006}
{Croton} D.~J.  et~al., 2006, MNRAS, 365, 11

\bibitem[Davis \& Laor(2011)]{dav11} 
Davis, S.~W., \& Laor, A.\ 2011, \apj, 728, 98 

\bibitem[Delvecchio et al.(2014)]{del14} 
Delvecchio, I., Gruppioni, C., Pozzi, F., et al.\ 2014, \mnras, 439, 2736 

\bibitem[Devecchi \& Volonteri(2009)]{dev09} 
Devecchi, B., \& Volonteri, M.\ 2009, \apj, 694, 302 

\bibitem[\protect\citeauthoryear{{Di Matteo}, {Colberg}, {Springel},
  {Hernquist} \& {Sijacki}}{{Di Matteo} et~al.}{2008}]{Dimatteo2008}
{Di Matteo} T.,  {Colberg} J.,  {Springel} V.,  {Hernquist} L.,    {Sijacki}
  D.,  2008, ApJ, 676, 33

\bibitem[\protect\citeauthoryear{{Dubois}, {Devriendt}, {Slyz} \&
  {Teyssier}}{{Dubois} et~al.}{2010}]{duboisetal10}
{Dubois} Y.,  {Devriendt} J.,  {Slyz} A.,    {Teyssier} R.,  2010, \mnras, 409,
  985

\bibitem[\protect\citeauthoryear{{Dubois}, {Devriendt}, {Slyz} \&
  {Teyssier}}{{Dubois} et~al.}{2012}]{2012MNRAS.420.2662D}
{Dubois} Y.,  {Devriendt} J.,  {Slyz} A.,    {Teyssier} R.,  2012, \mnras, 420,
  2662

\bibitem[Elvis et al.(2002)]{elv02} 
Elvis, M., Risaliti, G., \& Zamorani, G.\ 2002, \apjl, 565, L75 

\bibitem[Fanidakis et al.(2011)]{fan11} 
Fanidakis, N., Baugh, C.~M., Benson, A.~J., et al.\ 2011, \mnras, 410, 53 

\bibitem[Fanidakis et al.(2012)]{fan12} 
Fanidakis, N., Baugh, C.~M., Benson, A.~J., et al.\ 2012, \mnras, 419, 2797 

\bibitem[\protect\citeauthoryear{{Fontanot}, {Pasquali}, {De Lucia}, {van den
  Bosch}, {Somerville} \& {Kang}}{{Fontanot} et~al.}{2011}]{Fontanot2011}
{Fontanot} F.,  {Pasquali} A.,  {De Lucia} G.,  {van den Bosch} F.~C.,
  {Somerville} R.~S.,    {Kang} X.,  2011, \mnras, 413, 957

\bibitem[Goulding et al.(2010)]{gou10} 
Goulding, A.~D., Alexander, D.~M., Lehmer, B.~D., \& Mullaney, J.~R.\
2010, \mnras, 406, 597

\bibitem[Graham(2007)]{gra07} 
Graham, A.~W.\ 2007, \mnras, 379, 711 

\bibitem[Granato et al.(2001)]{gra01} 
Granato, G.~L., Silva, L., Monaco, P., et al.\ 2001, \mnras, 324, 757 

\bibitem[Greene \& Ho(2007)]{gre07} 
Greene, J.~E., \& Ho, L.~C.\ 2007, \apj, 667, 131 

\bibitem[Haiman 
\& Loeb(1998)]{1998ApJ...503..505H} Haiman, Z., \& Loeb, A.\ 1998, \apj, 503, 505 

\bibitem[Haiman et al.(2004)]{2004ApJ...612..698H} Haiman, Z., Quataert, 
E., \& Bower, G.~C.\ 2004, \apj, 612, 698 

\bibitem[H{\"a}ring \& Rix(2004)]{har04} 
H{\"a}ring, N., \& Rix, H.-W.\ 2004, \apjl, 604, L89 

\bibitem[\protect\citeauthoryear{{Haehnelt} \& {Kauffmann}}{{Haehnelt} \&
  {Kauffmann}}{2000}]{2000MNRAS.318L..35H}
{Haehnelt} M.~G.,  {Kauffmann} G.,  2000, \mnras, 318, L35

\bibitem[Hasinger(2008)]{has08} 
Hasinger, G.\ 2008, \aap, 490, 905 

\bibitem[Heckman et al.(2004)]{hec04} 
Heckman, T.~M., Kauffmann, G., Brinchmann, J., et al.\ 2004, \apj, 613, 109 

\bibitem[\protect\citeauthoryear{{Hirschmann}, {Somerville}, {Naab} \&
  {Burkert}}{{Hirschmann} et~al.}{2012}]{Hirschmann2012}
{Hirschmann} M.,  {Somerville} R.~S.,  {Naab} T.,    {Burkert} A.,  2012,
  \mnras, 426, 237
    
\bibitem[Hirschmann et al.(2014)]{2014MNRAS.442.2304H} Hirschmann, M., 
Dolag, K., Saro, A., et al.\ 2014, \mnras, 442, 2304 

\bibitem[Hopkins \& Hernquist(2009)]{hop09} 
Hopkins, P.~F., \& Hernquist, L.\ 2009, \apj, 698, 1550 

\bibitem[Hopkins et al.(2006)]{hop06} 
Hopkins, P.~F., Narayan, R., \& Hernquist, L.\ 2006, \apj, 643, 641 

\bibitem[Hopkins et al.(2007)]{hop07} 
Hopkins, P.~F., Richards, G.~T., \& Hernquist, L.\ 2007, \apj, 654, 731 

\bibitem[\protect\citeauthoryear{{Kauffmann} \& {Haehnelt}}{{Kauffmann} \&
  {Haehnelt}}{2000}]{Kauffmann2000}
{Kauffmann} G.,  {Haehnelt} M.,  2000, MNRAS, 311, 576

\bibitem[Kauffmann \& Heckman(2009)]{kau09} 
Kauffmann, G., \& Heckman, T.~M.\ 2009, \mnras, 397, 135 

\bibitem[Kelly \& Merloni(2012)]{kel12} 
Kelly, B.~C., \& Merloni, A.\ 2012, Advances in Astronomy, 2012, 970858 

\bibitem[Kelly \& Shen(2013)]{kel13} 
Kelly, B.~C., \& Shen, Y.\ 2013, \apj, 764, 45 

\bibitem[Kelly et al.(2009)]{kel09} 
Kelly, B.~C., Vestergaard, M., \& Fan, X.\ 2009, \apj, 692, 1388 

\bibitem[Kormendy 
\& Richstone(1995)]{1995ARA&A..33..581K} Kormendy, J., \& Richstone, D.\ 1995, \araa, 33, 581 

\bibitem[Kormendy \& Ho(2013)]{kor13} 
Kormendy, J., \& Ho, L.~C.\ 2013, \araa, 51, 511

\bibitem[Koushiappas et al.(2004)]{kou04} 
Koushiappas, S.~M., Bullock, J.~S., \& Dekel, A.\ 2004, \mnras, 354, 292 

\bibitem[Iwasawa et al.(2012)]{iwa12} 
Iwasawa, K., Gilli, R., Vignali, C., et al.\ 2012, \aap, 546, A84 

\bibitem[La Franca et al.(2005)]{laf05} 
La Franca, F., Fiore, F., Comastri, A., et al.\ 2005, \apj, 635, 864 

\bibitem[Li et al.(2011)]{li11} 
Li, Y.-R., Ho, L.~C., \& Wang, J.-M.\ 2011, \apj, 742, 33 

\bibitem[Li et al.(2012)]{li12} 
Li, Y.-R., Wang, J.-M., \& Ho, L.~C.\ 2012, \apj, 749, 187 

\bibitem[Lodato \& Natarajan(2007)]{lod07} 
Lodato, G., \& Natarajan, P.\ 2007, \mnras, 377, L64 

\bibitem[Lusso et al.(2012)]{lus12} 
Lusso, E., Comastri, A., Simmons, B.~D., et al.\ 2012, \mnras, 425, 623 

\bibitem[McConnell \& Ma(2013)]{mcc13} 
McConnell, N.~J., \& Ma, C.-P.\ 2013, \apj, 764, 184 

\bibitem[Marconi et al.(2004)]{mar04} 
Marconi, A., Risaliti, G., Gilli, R., et al.\ 2004, \mnras, 351, 169 

\bibitem[Merloni \& Heinz(2008)]{mer08} 
Merloni, A., \& Heinz, S.\ 2008, \mnras, 388, 1011 

\bibitem[Merloni et al.(2014)]{mer14} 
Merloni, A., Bongiorno, A., Brusa, M., et al.\ 2014, \mnras, 437, 3550 

\bibitem[\protect\citeauthoryear{{Monaco}, {Fontanot} \& {Taffoni}}{{Monaco}
  et~al.}{2007}]{2007MNRAS.375.1189M}
{Monaco} P.,  {Fontanot} F.,    {Taffoni} G.,  2007, \mnras, 375, 1189

\bibitem[\protect\citeauthoryear{{Monaco}, {Salucci} \& {Danese}}{{Monaco}
  et~al.}{2000}]{2000MNRAS.311..279M}
{Monaco} P.,  {Salucci} P.,    {Danese} L.,  2000, \mnras, 311, 279

\bibitem[Nobuta et al.(2012)]{nob12} 
Nobuta, K., Akiyama, M., Ueda, Y., et al.\ 2012, \apj, 761, 143 

\bibitem[Panessa et al.(2006)]{pan06} 
Panessa, F., Bassani, L., Cappi, M., et al.\ 2006, \aap, 455, 173 

\bibitem[Raimundo et al.(2012)]{rai12} 
Raimundo, S.~I., Fabian, A.~C., Vasudevan, R.~V., Gandhi, P., 
\& Wu, J.\ 2012, \mnras, 419, 2529 

\bibitem[Schulze \& Wisotzki(2010)]{sch10} 
Schulze, A., \& Wisotzki, L.\ 2010, \aap, 516, A87

\bibitem[Schulze et al.(2015)]{sch15} 
Schulze, A., Bongiorno, A., Gavignaud, I., et al.\ 2015, \mnras, 447, 2085 

\bibitem[Shakura \& Sunyaev(1973)]{sha73} 
Shakura, N.~I., \& Sunyaev, R.~A.\ 1973, \aap, 24, 337 

\bibitem[Shankar(2013)]{sha13b} 
Shankar, F.\ 2013, Classical and Quantum Gravity, 30, 244001 

\bibitem[Shankar et al.(2009)]{sha09} 
Shankar, F., Weinberg, D.~H., \& Miralda-Escud{\'e}, J.\ 2009, \apj, 690, 20 

\bibitem[Shankar et al.(2013)]{sha13} 
Shankar, F., Weinberg, D.~H., \& Miralda-Escud{\'e}, J.\ 2013, \mnras,
428, 421

\bibitem[Shankar et al.(2010)]{2010MNRAS.401.1869S} Shankar, F., Sivakoff, 
G.~R., Vestergaard, M., \& Dai, X.\ 2010, \mnras, 401, 1869 

\bibitem[Shen \& Kelly(2012)]{she12} 
Shen, Y., \& Kelly, B.~C.\ 2012, \apj, 746, 169 

\bibitem[\protect\citeauthoryear{{Sijacki}, {Springel}, {di Matteo} \&
  {Hernquist}}{{Sijacki} et~al.}{2007}]{Sijacki2007}
{Sijacki} D.,  {Springel} V.,  {di Matteo} T.,    {Hernquist} L.,  2007, MNRAS,
  380, 877

\bibitem[\protect\citeauthoryear{{Sijacki}, {Vogelsberger}, {Genel},
  {Springel}, {Torrey}, {Snyder}, {Nelson} \& {Hernquist}}{{Sijacki}
  et~al.}{2015}]{2015MNRAS.452..575S}
{Sijacki} D.,  {Vogelsberger} M.,  {Genel} S.,  {Springel} V.,  {Torrey} P.,
  {Snyder} G.~F.,  {Nelson} D.,    {Hernquist} L.,  2015, \mnras, 452, 575

\bibitem[Small \& Blandford(1992)]{sma92} 
Small, T.~A., \& Blandford, R.~D.\ 1992, \mnras, 259, 725 

\bibitem[Soltan(1982)]{sol82} 
Soltan, A.\ 1982, \mnras, 200, 115 

\bibitem[Trump et al.(2011)]{tru11} 
Trump, J.~R., Impey, C.~D., Kelly, B.~C., et al.\ 2011, \apj, 733, 60 

\bibitem[Ueda et al.(2003)]{ued03} 
Ueda, Y., Akiyama, M., Ohta, K., \& Miyaji, T.\ 2003, \apj, 598, 886 

\bibitem[Ueda et al.(2014)]{ued14} 
Ueda, Y., Akiyama, M., Hasinger, G., Miyaji, T., \& Watson, M.~G.\ 
2014, \apj, 786, 104 

\bibitem[Veale et al.(2014)]{vea14} 
Veale, M., White, M., \& Conroy, C.\ 2014, \mnras, 445, 1144 

\bibitem[Vestergaard \& Osmer(2009)]{ves09} 
Vestergaard, M., \& Osmer, P.~S.\ 2009, \apj, 699, 800 

\bibitem[Vestergaard et al.(2008)]{ves08} 
Vestergaard, M., Fan, X., Tremonti, C.~A., Osmer, P.~S., 
\& Richards, G.~T.\ 2008, \apjl, 674, L1 

\bibitem[Vika et al.(2009)]{vik09} 
Vika, M., Driver, S.~P., Graham, A.~W., \& Liske, J.\ 2009, \mnras, 400, 1451 

\bibitem[\protect\citeauthoryear{{Volonteri}, {Haardt} \& {Madau}}{{Volonteri}
  et~al.}{2003}]{VHM}
{Volonteri} M.,  {Haardt} F.,    {Madau} P.,  2003, ApJ, 582, 559

\bibitem[Volonteri(2010)]{vol10} 
Volonteri, M.\ 2010, \aapr, 18, 279 

\bibitem[Volonteri et al.(2007)]{vol07} 
Volonteri, M., Sikora, M., \& Lasota, J.-P.\ 2007, \apj, 667, 704 

\bibitem[Volonteri et al.(2008)]{vol08} 
Volonteri, M., Lodato, G., \& Natarajan, P.\ 2008, \mnras, 383, 1079 

\bibitem[Volonteri et al.(2005)]{vol05} 
Volonteri, M., Madau, P., Quataert, E., \& Rees, M.~J.\ 2005, \apj, 620, 69 

\bibitem[Volonteri et al.(2016)]{2016arXiv160201941V} Volonteri, M., 
Dubois, Y., Pichon, C., \& Devriendt, J.\ 2016, arXiv:1602.01941 

\bibitem[Yu \& Tremaine(2002)]{yu02} 
Yu, Q., \& Tremaine, S.\ 2002, \mnras, 335, 965 

\bibitem[Yu et al.(2005)]{yu05} 
Yu, Q., Lu, Y., \& Kauffmann, G.\ 2005, \apj, 634, 901 

\bibitem[Wang et al.(2006)]{wan06} 
Wang, J.-M., Chen, Y.-M., \& Zhang, F.\ 2006, \apjl, 647, L17 

\bibitem[Wang et al.(2009)]{wan09} 
Wang, J.-M., Hu, C., Li, Y.-R., et al.\ 2009, \apjl, 697, L141 

\bibitem[Wu et al.(2013)]{wu13} 
Wu, S., Lu, Y., Zhang, F., \& Lu, Y.\ 2013, \mnras, 436, 3271

\bibitem[Wyithe 
\& Loeb(2002)]{2002ApJ...581..886W} Wyithe, J.~S.~B., \& Loeb, A.\ 2002, \apj, 581, 886 

\bibitem[Wyithe 
\& Loeb(2003)]{2003ApJ...595..614W} Wyithe, J.~S.~B., \& Loeb, A.\ 2003, \apj, 595, 614 

\end{thebibliography}

\begin{appendix}

\section{Testing model assumptions}
\label{app1}

\begin{figure}
  \centering
  \includegraphics[width=\hsize]{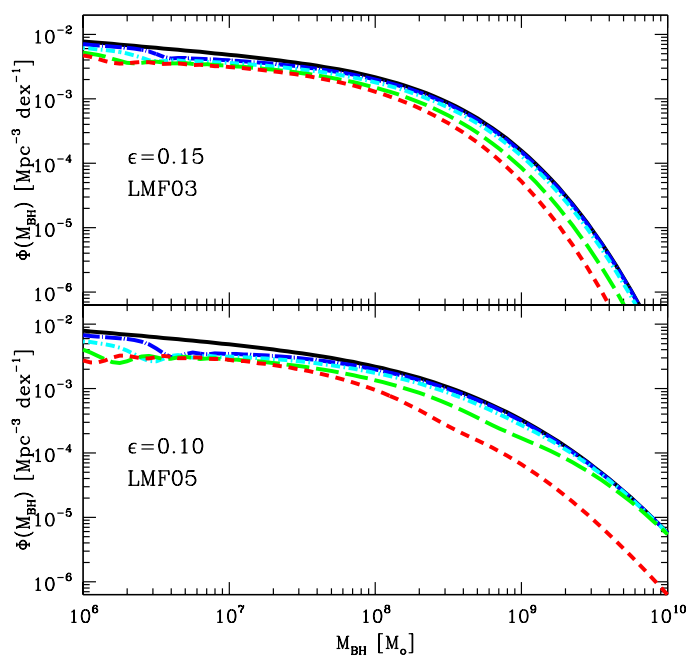}
  \caption{SMBH mass function obtained with $\epsilon=0.15$ and the
    LMF03 local BHMF (upper panel), and with $\epsilon=0.1$ and the
    LMF05 local BHMF (lower panel). The meaning of the lines are like
    in Fig.\,\ref{s4f2}.}
  \label{s4f4a}
\end{figure}

\begin{figure}
  \centering
 \includegraphics[width=\hsize]{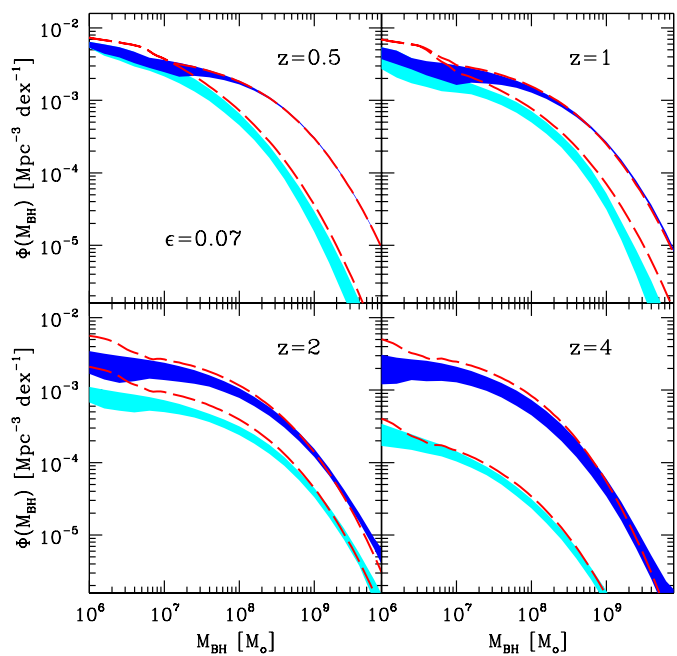}
 \caption{Uncertainties on the mass function of SMBHs (shaded blue
   areas) and of AGN (shaded cyan area) predicted by the model with
   $\epsilon=0.07$. Dashed red lines are the results using
   $\lambda_{cut}=10^{-5}$.}
  \label{a1f1}
\end{figure}

Here we discuss solutions of the continuity equation with different
values/combinations of the radiative efficiency and the local
BHMF. Firstly, let us consider models with a local BHMF obtained by
the convolution with 0.3\,dex Gaussian scatter (LMF03). We only find
acceptable fits to the observational QLF for
$\epsilon\ga0.1$. The results for $\epsilon=0.1$ were discussed in the
main part of the paper. When $\epsilon>0.1$, the BHMF does not evolve
significantly over time for $M\la10^8\msol$, as shown by
Fig.\,\ref{s4f4a}. Instead, for $\epsilon<0.1,$ the evolution of
high--mass SMBHs is too fast and, consequently, the model strongly
underestimates the high-luminosity part of the observational QLF at
high redshifts because of the lack of active SMBHs with
$M\gg10^8\msol$. On the other hand, if the local mass function with
0.5\,dex Gaussian scatter (LMF05) is used, we typically find better
fits to the QLF independent of the choice of the radiative
efficiency. Nevertheless, again, when $\epsilon\ga0.1$, the BHMF
exhibits a smaller evolution over the whole range of masses (see
Fig.\,\ref{s4f4a}).

These results can be understood taking into account that the average
accretion rate is proportional to $(1-\epsilon)/\epsilon$ and
increases by a factor 2 reducing $\epsilon$ from 0.1 to 0.05.  If we
use a local mass function with low number density of very massive
SMBHs (as the LMF03 one), the duty cycle at these masses will be large
in order to fit the high-luminosity tail of the QLF. The combination
of a low radiative efficiency and a high duty cycle gives a high
accretion rate and a fast evolution of high-mass SMBHs, whose
number density will become too small at high redshifts. Viceversa, the
argument can be overturned in the case of high radiative efficiencies
associated to the LMF05 local mass function: the non-evolving MFs
discussed above are then observed. We can conclude that,
independently of the local BHMF, models with $\epsilon>0.1$ seem to be
disfavored. They entail, in fact, an average accretion rate that is too
low to produce SMBHs significantly growing with redshifts.

Fig.\,\ref{a1f1} shows the 1-$\sigma$ errors associated to the SMBH
and AGN MFs for $\epsilon=0.07$, obtained from the MCMC method; they
are relatively small, due to the strong constraints imposed by the quasar
luminosity function. Clearly, they are not representative of the
uncertainties of the model, that are mainly drawn by the uncertainties
on the value of the average radiative efficiency and on the Eddington
ratio distribution.

In Fig.\,\ref{a1f1}, we also investigate the effect of varying
$\lambda_{cut}$, which defines the minimum accretion rate for an
active SMBH. We extrapolate the power-law behavior of the Eddington
ratio distribution for type-2 AGN up to $\lambda_{cut}=10^{-5}$
(type-1 AGN are not affected by the choice of $\lambda_{cut}$). This
is probably an extreme situation because we expect a drop in
$P(\lambda)$ somewhere below $\lambda\simeq10^{-4}$
\citep{hop09,fan12}. As expected, we find larger MFs, but the effect
is modest both for SMBHs and for AGN.

\begin{figure}
  \centering
  \includegraphics[width=9cm]{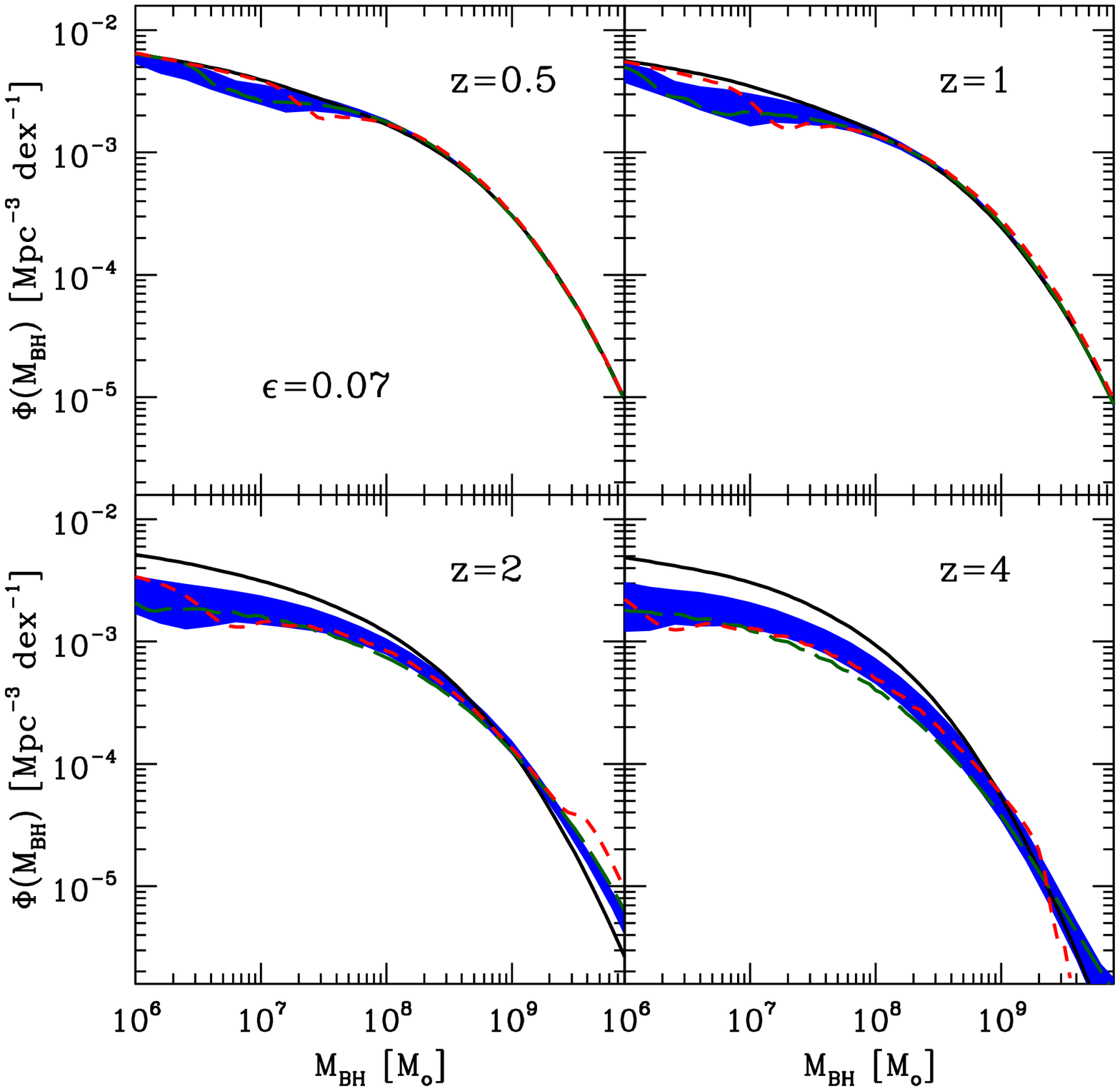}
  \includegraphics[width=9cm]{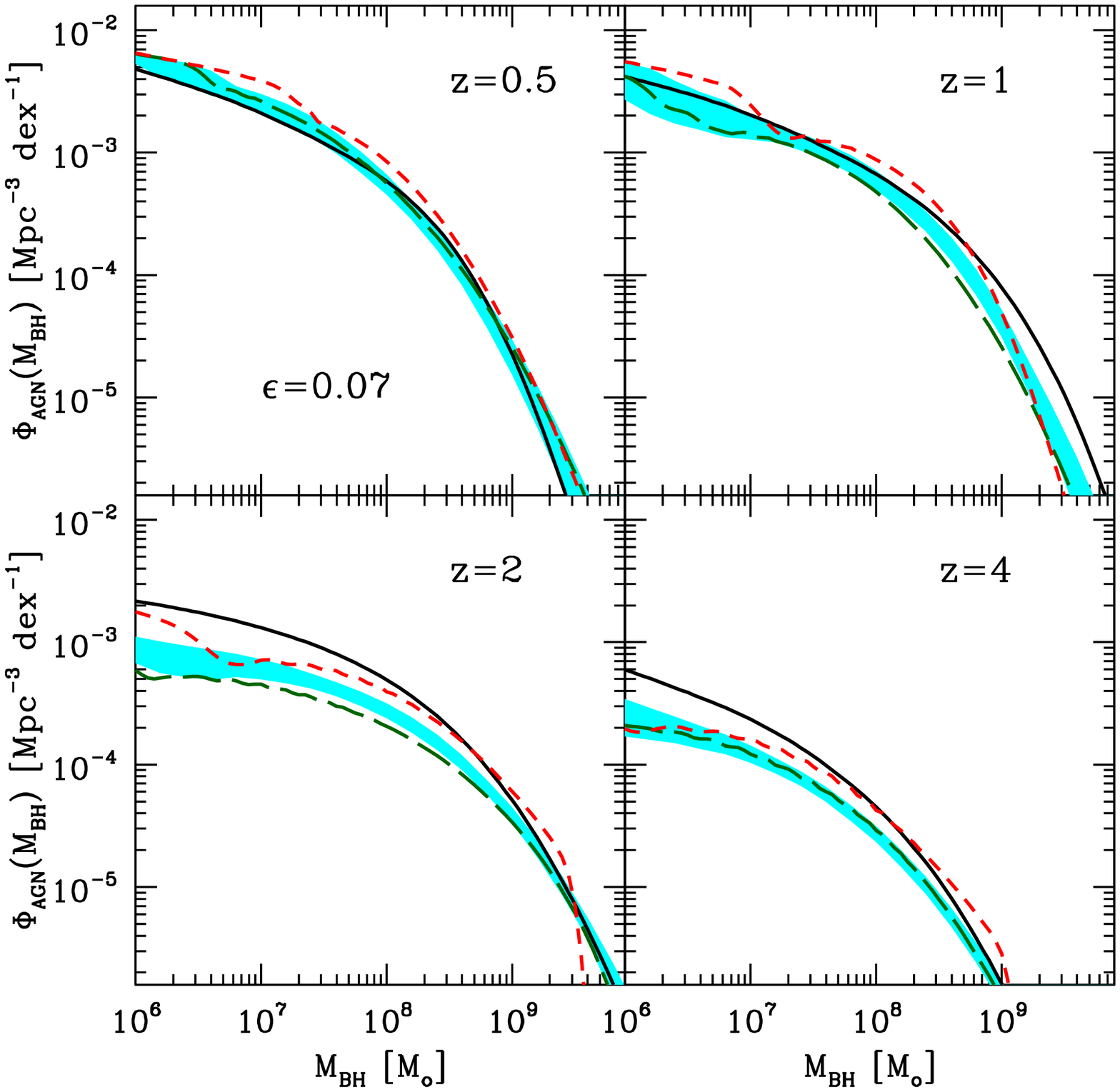}
  \caption{Mass function of SMBHs (top panel) and AGN (bottom panel)
    obtained after changing the Eddington ratio distribution. Lines
    are for following cases: (1) we use a truncated power law for
    $P(\lambda)$ of type-1 AGN (red short dashed lines); (2) a
    power-law distribution for type-2 AGN with constant slope
    $\alpha_{\lambda}=-0.6$ (black solid lines); (3) a faster evolving
    power-law distribution ($\alpha_{\lambda}=-0.6/(0.4+z)^{1.5}$)
    for type-2 AGN (green long dashed lines). Shaded areas are for
    the uncertainty of the model as in Fig.\,\ref{a1f1}.}
 \label{a1f2}
\end{figure}

Observations give only very partial constraints on the Eddington ratio
distribution. For type-1 AGN they are limited to the highest values
of the Eddington ratio ($\lambda\ga0.1$), while for type-2 AGN the
Eddington ratio distribution is determined over a large range of
accretion rates (from $\sim10^{-4}$ to 1) but only for $z<1$. Below, we discuss how the model predictions (especially in relation to
the AGN MF) can change if different Eddington ratio distributions are
employed. Results are shown in Fig.\,\ref{a1f2}.

\begin{itemize}

\item So far we have used a log-normal function to fit the Eddington
  ratio distributions estimated by \citet{kel13}. However, this is not
  the only possible choice. A truncated power law, as used for type-2
  AGN (Eq.\,\ref{s3e3}), also provides a good fit to data. We use the
  same power-law index as for type-2 AGN (Eq.\,\ref{s3e3b}) and we
  find the cut-off $\lambda_0$ that best fits the \citet{kel13}
  estimates at the different redshift bins. BHMFs are well compatible
  with previous results, while AGN MFs are typically larger,
  especially at high redshifts.

\item The Eddington ratio distribution for type-2 AGN is determined
  only for $z<1$. We therefore test different redshift dependences of
  the power-law index with respect to the one used in
  Eq.\,\ref{s3e3b}. We consider two opposite situations: (1) we
  extrapolate the distribution observed at low redshifts to high
  redshifts, keeping a constant slope $\alpha_{\lambda}=-0.6$ over the
  whole redshift range; or (2) we use a steeper redshift dependence,
  $\alpha_{\lambda}=-0.6/(0.4+z)^{1.5}$. As expected, the former gives
  mass functions that evolve more slowly at $z\ga1$. The effect is
  observed mainly at low/intermediate masses, at which the number
  density is larger by a factor $\la2$ with respect to the reference
  model. Finally, the second model does not produce significant
  changes in MFs.

\end{itemize}

We can conclude that using different Eddington ratio distributions
seems to have a limited impact on the BHMF evolution. The uncertainty
on this is dominated by the radiative efficiency, that can vary from
$\epsilon=0.05$ to 0.1. In principle, the AGN MF should be more
sensitive to the choice of $P(\lambda)$. However, we see that the
changes are not particularly relevant; they are only significantly larger than the
intrinsic uncertainty associated to the models when we take a
constant Eddington ratio distribution for type-2 AGN. The uncertainty
on the Eddington ratio distribution then simply translates into a
small increase of the uncertainty on the model predictions. This
proves that our approach gives robust constraints on the AGN MF up to
redshifts 4.

\end{appendix}

\end{document}